\documentclass[prodmode,acmtecs,acmsmall]{acmart} 

\acmVolume{0}
\acmNumber{0}
\acmArticle{0}
\acmYear{0000}
\acmMonth{0}

\usepackage{amsmath,amsthm}
\usepackage[UKenglish]{babel}
\usepackage{enumitem}
\setlist{itemsep=0pt, topsep=0.5em}
\setlist[description]{leftmargin=*, itemsep = 5pt}
\usepackage{tikz-cd}
\usepackage{tikz}
\usetikzlibrary{arrows.meta}
\usepackage{mathtools}
\usepackage{graphicx}
\usepackage{mathrsfs}
\usepackage{stmaryrd}
\usepackage[scr=euler]{mathalfa}
\usepackage{hyperref}
\hypersetup{
breaklinks=true, 
hidelinks,
bookmarks=true,     
bookmarksopen=true, 
  }
\usepackage{nicematrix}
\usepackage{pifont}
\providecommand{\noopsort}[1]{}

\AtEndDocument{%
  \addtocontents{toc}{\protect\setcounter{tocdepth}{3}}
}

\providecommand{\noopsort}[1]{}

\usepackage{scalerel}
\DeclareFontFamily{U}{mathb}{\hyphenchar\font45}
\DeclareFontShape{U}{mathb}{m}{n}{
      <5> <6> <7> <8> <9> <10> gen * mathb
      <10.95> mathb10 <12> <14.4> <17.28> <20.74> <24.88> mathb12
      }{}
\DeclareSymbolFont{mathb}{U}{mathb}{m}{n}
\DeclareMathSymbol{\boxslash}{2}{mathb}{"6D}

\newcommand{\Fraisse}{Fra\"{i}ss\'{e}}
\newcommand{\EF}{Ehrenfeucht--\Fraisse}
\renewcommand{\epsilon}{\varepsilon}
\renewcommand{\theta}{\vartheta}
\renewcommand{\phi}{\varphi}

\newcommand{\N}{\mathbb{N}}

\newcommand{\down}{{\downarrow}\,} 
\newcommand{\cvr}{\prec} 
\DeclareMathOperator{\htf}{\mathrm{ht}} 
\newcommand{\llp}[1]{{}^{\scaleobj{0.7}{\boxslash}}#1} 
\newcommand{\rlp}[1]{#1^{\mkern1mu \scaleobj{0.7}{\boxslash}}} 

\newcommand{\w}{\widehat}
\newcommand{\op}{\mathrm{op}} 
\DeclareMathOperator{\dom}{dom} 

\newcommand{\sg}{\sigma}

\newcommand{\tr}[1]{\llbracket #1 \rrbracket}
\DeclareMathOperator{\tp}{tp}
\newcommand{\n}{\mathbf{n}}
\newcommand{\IMP}{\Rrightarrow}

\newcommand{\W}{\mathscr{W}} 
\newcommand{\Ga}{\mathscr{G}} 
\newcommand{\one}{\mathbf{1}} 
\newcommand{\ud}{\dot} 
\newcommand{\acc}{\frown} 

\newcommand{\G}{\mathbb{G}} 
\newcommand{\Ek}{\mathbb{E}_k} 
\newcommand{\EkI}{\mathbb{E}_k^{I}} 
\newcommand{\Eo}{\mathbb{E}_{\omega}} 
\newcommand{\Mk}{\mathbb{M}_k} 
\newcommand{\Mo}{\mathbb{M}_{\omega}} 
\newcommand{\Pn}{\mathbb{P}_n} 
\newcommand{\PnI}{\mathbb{P}_n^{I}} 
\DeclareMathOperator{\Emb}{\mathbb{S}} 
\DeclareMathOperator{\Path}{\mathbb{P}} 
\newcommand{\Pathast}{\mathbb{P}^{*}} 

\newcommand{\FI}{\mathscr{F}^{I}} 

\DeclareMathOperator{\Set}{\mathbf{Set}} 
\DeclareMathOperator{\C}{\mathscr{C}} 
\DeclareMathOperator{\A}{\mathscr{A}} 
\DeclareMathOperator{\T}{\mathscr{T}} 
\DeclareMathOperator{\E}{\mathscr{E}} 
\newcommand{\CSstar}{\mathbf{Struct}_{\bullet}(\sg)} 
\newcommand{\CSplus}{\mathbf{Struct}(\sg^I)} 
\newcommand{\CS}{\mathbf{Struct}(\sg)}
\newcommand{\R}{\mathscr{R}} 

\newcommand{\KL}{\mathbf{Kl}} 
\newcommand{\EM}{\mathbf{EM}} 

\newcommand{\Ob}{\mathrm{Ob}} 
\DeclareMathOperator{\D}{\mathscr{D}} 
\newcommand{\St}{\mathbf{S}} 
\renewcommand{\P}{\mathbf{P}} 
\newcommand{\Ext}{\mathbf{Ext}} 

\newcommand{\LL}{\mathcal{L}} 
\newcommand{\FO}{\mathrm{FO}} 
\newcommand{\EPFO}{\exists^+\mathrm{FO}} 
\newcommand{\EFO}{\exists\mathrm{FO}} 
\newcommand{\ML}{\mathrm{ML}} 
\newcommand{\EML}{\exists\mathrm{ML}} 
\newcommand{\EPML}{\exists^{+}\mathrm{ML}} 

\newcommand{\emb}{\rightarrowtail} 
\newcommand{\epi}{\twoheadrightarrow} 
\newcommand{\rl}{\rightleftarrows}
\newcommand{\lr}{\mathrel{\raisebox{1.5pt}{$\underline{\leftrightarrow}$}}}
\newcommand{\pb}{\mathrel{\raisebox{1.5pt}{$\underline{\rightarrow}$}}}
\newcommand{\id}{\mathrm{id}} 
\newcommand{\Q}{\mathscr{Q}} 
\newcommand{\M}{\mathscr{M}} 
\newcommand{\Cof}{\mathsf{C}} 
\newcommand{\We}{\mathsf{W}} 
\newcommand{\Fib}{\mathsf{F}} 
\DeclareMathOperator{\cof}{\tikz[baseline=-0.6ex]\draw[{Triangle[reversed][length=3pt]}-{Classical TikZ Rightarrow[length=2pt]}] (0,0) -- +(0.5,0);} 
\DeclareMathOperator{\bwe}{\tikz[baseline=-0.6ex]\draw[-{Classical TikZ Rightarrow[length=2pt]}] (0,0) -- +(0.6,0);} 
\DeclareMathOperator{\fib}{\tikz[baseline=-0.6ex]\draw[-{Triangle[length=3pt]}] (0,0) -- +(0.5,0);} 
\DeclareMathOperator{\longfib}{\tikz[baseline=-0.6ex]\draw[-{Triangle[length=3pt]}] (0,0) -- +(0.6,0);} 
\DeclareMathOperator{\longleftfib}{\tikz[baseline=-0.6ex]\draw[{Triangle[length=3pt]}-] (0,0) -- +(0.65,0);} 

\makeatletter
\newcommand{\we}{}
\DeclareRobustCommand{\we}{%
  \mathrel{\vphantom{\rightarrow}\mathpalette\filledcircle@arrow\relax}%
}
\newcommand{\filledcircle@arrow}[2]{%
  \m@th
  \ooalign{%
    \hidewidth$#1\bullet\mkern1mu$\hidewidth\cr
    $#1\bwe$\cr}%
}
\makeatother

\makeatletter
\newcommand{\tfib}{}
\DeclareRobustCommand{\tfib}{%
  \mathrel{\vphantom{\fib}\mathpalette\filledcircle@fib\relax}%
}
\newcommand{\filledcircle@fib}[2]{%
  \m@th
  \ooalign{%
    \hidewidth$#1\bullet\mkern1mu$\hidewidth\cr
    $#1\longfib$\cr}%
}

\newcommand{\tfibleft}{}
\DeclareRobustCommand{\tfibleft}{%
  \mathrel{\vphantom{\fib}\mathpalette\filledcircle@twoheadleftarrow\relax}%
}
\newcommand{\filledcircle@twoheadleftarrow}[2]{%
  \m@th
  \ooalign{%
    \hidewidth$#1\bullet$\hidewidth\cr
    $#1\longleftfib\mkern1mu$\cr}%
}
\makeatother

\usepackage[cmtip,all]{xy}
\newcommand{\longsquiggly}{\xymatrix{{}\ar@{~>}[r]&{}}}

\tikzcdset{
    relay arrow/.default = 10pt,
    relay arrow/.style = {
        rounded corners,
        to path = {
            \pgfextra{
                \def\sourcecoordinate{\pgfpointanchor{\tikztostart}{center}}
                \def\targetcoordinate{\pgfpointanchor{\tikztotarget}{center}}
                \pgfmathanglebetweenpoints{\sourcecoordinate}{\targetcoordinate}
                \edef\tempangle{\pgfmathresult}
                \pgftransformrotate{\tempangle}
                \pgfmathifthenelse{#1>0}{\tempangle+90}{\tempangle-90}
                \pgfcoordinate{tempcoord}{\pgfpointanchor{\tikztostart}{\pgfmathresult}}
            }
            (tempcoord)
            -- ([yshift=#1]tempcoord)
            -- ([yshift=#1]tempcoord-|\tikztotarget.center)\tikztonodes
            --(\tikztotarget)
        }
    }
}

\usetikzlibrary{calc}
\tikzset{hbend left/.style={
    to path={let \p1=($(\tikztotarget)-(\tikztostart)$),
        \p2=($(\tikztostart.north)-(\tikztostart.south)$),
        \p3=($(\tikztotarget.north)-(\tikztotarget.south)$),
        \n1={max(\y2,\y3)/2} in
    \ifdim\x1>0pt
     ([yshift=\n1]\tikztostart.east) to[bend left] ([yshift=\n1]\tikztotarget.west)
    \else
     ([yshift=-0.7*\n1]\tikztostart.west) to[bend left] ([yshift=-0.7*\n1]\tikztotarget.east)
    \fi
    }}}

\makeatletter
\newcommand{\open}{}
\DeclareRobustCommand{\open}{%
  \mathrel{\vphantom{\rightarrow}\mathpalette\circle@arrow\relax}%
}
\newcommand{\circle@arrow}[2]{%
  \m@th
  \ooalign{%
    \hidewidth$#1\circ\mkern1mu$\hidewidth\cr
    $#1\longrightarrow$\cr}%
}

\newcommand{\openleft}{}
\DeclareRobustCommand{\openleft}{%
  \mathrel{\vphantom{\leftarrow}\mathpalette\circle@leftarrow\relax}%
}
\newcommand{\circle@leftarrow}[2]{%
  \m@th
  \ooalign{%
    \hidewidth$#1\circ$\hidewidth\cr
    $#1\longleftarrow\mkern1mu$\cr}%
}
\makeatother

\usetikzlibrary{decorations.markings}
\tikzset{circlearrow/.style={/utils/exec=\tikzset{every node/.append style={outer sep=0.6ex}},
postaction=decorate,decoration={markings,
mark=at position 0.5 with {\draw circle [radius=1.75pt];}}},
circlearrow/.default=0.75ex}

\begin{document}

\theoremstyle{acmdefinition}
\newtheorem{remark}[theorem]{Remark}


\title{An invitation to game comonads}

\author{Samson Abramsky}
\author{Luca Reggio}
\affiliation{
\institution{Department of Computer Science, University College London}
\city{66--72 Gower Street, London, WC1E 6EA}
\country{United Kingdom}
}

\addtocontents{toc}{\protect\setcounter{tocdepth}{0}}
\begin{abstract}
Game comonads offer a categorical view of a number of model-comparison games central to  model theory, such as pebble and \EF~games. Remarkably, the categories of coalgebras for these comonads capture preservation of several fragments of resource-bounded logics, such as (infinitary) first-order logic with $n$ variables or bounded quantifier rank, and corresponding combinatorial parameters  such as tree-width and tree-depth. In this way, game comonads provide a new bridge between categorical methods developed for semantics, and the combinatorial and algorithmic methods of resource-sensitive model theory.

We give an overview of this framework and outline some of its applications, including the study of homomorphism counting results in finite model theory, and of equi-resource homomorphism preservation theorems in logic using the axiomatic setting of arboreal categories. Finally, we describe some homotopical ideas that arise naturally in the context of game comonads.
\end{abstract}

\maketitle

\tableofcontents
\addtocontents{toc}{\protect\setcounter{tocdepth}{1}}

\vspace{-2em}

\section{Introduction}

The traditional divide in theoretical computer science between formal methods and semantics on the one hand, and algorithms and computational complexity on the other, runs deep, as can be seen e.g.\ in the division into two volumes of the \emph{Handbook of theoretical computer science} \cite{handbooktcsA,handbooktcsB}. The first strand is exemplified by categorical semantics, while the second strand is exemplified by finite model theory and descriptive complexity. The aim of this survey article is to introduce the reader to the framework of game comonads, which bridges these two strands by bringing the tools of category theory to the field of finite and resource-sensitive model theory. This research programme was initiated in early 2017 and has been an active area of research since then.\footnote{As witnessed by the ever-increasing length of surveys on the subject!} Contributors to this programme include (in alphabetical order) Adam \'{O} Conghaile, Anuj Dawar, Tom\'a\v{s} Jakl, Thomas Laure, Dan Marsden, Yo\`{a}v Montacute, Tom Paine, Colin Riba, Nihil Shah and Pengming Wang.

\subsection{Finite and resource-sensitive model theory}
\textbf{Finite model theory} studies theories and their classes of \emph{finite} models, and has sparked researchers' interest since the early days of model theory, cf.\ e.g.\ \cite{Trakhtenbrot1950}. It has flourished and grown rapidly into an independent field in the 1970s--80s thanks to its connections with complexity theory and database theory. Nowadays, finite model theory is a subfield of mathematical logic which beautifully exemplifies the convergence of mathematics and computer science, see e.g.\ \cite{Libkin2004,GKLMSVVW2005}. 
Many results of classical model theory fail when restricted to finite models. These include G\"odel's Completeness Theorem, stating that a sentence is true in all models of a theory $T$ just when it can be proved from the axioms in~$T$, and the Compactness Theorem, stating that $T$ has a model just when every finite subset of~$T$ does. Model-theoretic tools can be used to study classes of finite models. This is the subject of \emph{pseudo-finite model theory} \cite{Chatzidakis1997,Hrushovski2002,Vaananen2003}, but this approach inevitably leads to considering infinite models. For example, the theory of finite fields (i.e., the collection of all sentences true in all finite fields) has infinite models, the pseudo-finite fields. 
On the other hand, finite model theory has developed its own methods, largely based on combinatorial and probabilistic techniques.

The deep link between finite model theory and complexity theory has its origin in Fagin's theorem, showing that the problems in the complexity class NP are exactly those definable in existential second-order logic \cite{Fagin1974}. This result initiated the subject of descriptive complexity, which classifies computational problems according to the logical language required to express them \cite{Immerman1999}. 
The idea of stratifying by \textbf{logical resources}, typically represented as complexity measures of formulas such as \emph{quantifier rank} or \emph{number of variables}, 
has had great success and is central to finite model theory, see e.g.\ \cite[Theorem~7.21 and Corollary~7.22]{Immerman1999} for a stratification by number of blocks of second-order quantifiers. It is reflected in the use of \textbf{model-comparison games}, such as \EF \ \cite{Ehrenfeucht1960,Fraisse1954} or pebble games \cite{Barwise1977,Immerman1982}, to determine if two models are equivalent in a given logic fragment. Here, the logical resources correspond to the resources available to the two players, Spoiler and Duplicator, such as the number of rounds or the number of pebbles (cf.\ \S\ref{s:games}). 

This method, wherein logical resources are placed centre stage, applies to both finite and infinite structures, leading to what we shall call \emph{resource-sensitive model theory}. It can also be thought of as ``model theory without compactness'', since it avoids the use of the Compactness Theorem. This viewpoint is at the heart of most preservation and inexpressibility results in finite model theory, and it has also led to remarkable resource-sensitive refinements of classical results such as the Equirank Homomorphism Preservation Theorem \cite{Rossman2008}; cf.\ \S\ref{s:HPT}.

\subsection{Game comonads}
In 2017, a novel approach to logical resources based on \textbf{game comonads}, providing categorical accounts of key constructions in finite model theory, was initiated in \cite{abramsky2017pebbling,DBLP:conf/csl/AbramskyS18}. 
The key insight is that model-comparison games can be naturally organised into endofunctors on categories of relational structures that carry the structure of comonads (these are the dual notion to monads, which are widely used in functional programming). 
Intuitively, given a type of game $G$ and a resource parameter~$k$, the game comonad~$\G_{k}$ assigns to a structure~$A$ a new structure~$\G_{k} A$ which can be thought of as a \emph{forest cover} (or forest-ordered decomposition) of $A$, abstracting the idea of \mbox{$k$-unravelling} from modal logic. The universe of~$\G_{k}A$ is the set of all possible plays in the structure $A$ in the game~$G$ in which the appropriate notion of resources is bounded by $k$. Thus, game comonads arise by viewing games not as external artefacts, but as semantic constructions in their own right. The first key observation is that Duplicator winning strategies for the existential $G$-game from $A$ to $B$ are recovered as Kleisli morphisms
\[
\G_{k}A\to B.
\] 
Duplicator winning strategies for the back-and-forth game, and variants thereof, can also be recovered.
In this manner, the notion of local approximation built into the game is internalised into the category of relational structures.

Remarkably, this idea applies to a number of games central to finite model theory, leading e.g.\ to \emph{\EF} and \emph{pebbling comonads}, corresponding to \EF~and pebble games, respectively.
In fact, game comonads cover a range of logics, including first-order, modal, guarded and hybrid logics, and logics with generalised quantifiers \cite{AS2021,conghaile2021game,Guarded2021,Hybrid2022}. See \cite{Abramsky2022} for a survey.
In each case, the categories of coalgebras for the comonads encode, in a syntax-free fashion, important logical resources and combinatorial parameters of structures; cf.\ Figure~\ref{figure:game-comonads}. 

\begin{figure}[h]
\centering
\resizebox{0.95\columnwidth}{!}{
\renewcommand{\arraystretch}{1.4}
\begin{NiceTabular}
   [
     columns-width=5cm,
     hvlines-except-borders,
     rules={color=white,width=1pt}
   ]
   {ccc}
\CodeBefore
  \rowcolor{cyan}{1}
  \rowcolors{2}{cyan!25}{cyan!15}
\Body
  \RowStyle[color=white]{}
   Game comonads      & Logical resources & Combinatorial parameters \\
  pebble comonad            & number of variables   &  tree-width    \\
  Ehrenfeucht--Fra\"iss\'e comonad            & quantifier rank      &  tree-depth    \\
  modal comonad            & modal depth   &  synchronization-tree depth \\
  guarded comonad            & guarded-quantifier depth    & $\mathfrak{g}$-guarded tree-width  \\
\end{NiceTabular}
}
\caption{Examples of game comonads, with corresponding logical resources and combinatorial parameters.}
\label{figure:game-comonads}
\end{figure}

Having established a common ground for various notions of logical resources in the form of game comonads and their coalgebras, the possibility opens up of treating logical resources uniformly, and comparing them by studying the structural properties of the corresponding game comonads. This can be done by working with \textbf{arboreal categories}, which provide an axiomatic approach to game comonads \cite{AR2021icalp,AR2022}. From this axiomatic perspective, interesting ``dividing lines'' can be identified, such as the \emph{bisimilar companion property} (see~\S\ref{s:HPT}).

\subsection{A guided bibliography}
In addition to the present one, surveys on game comonads include \cite{ABRAMSKY2020,Abramsky2022}; the first one is aimed more at theoretical computer scientists, the second at mathematical logicians. The ``first wave'' in this research programme consisted in establishing the paradigm and finding many examples:
\begin{itemize}
\item the first paper~\cite{abramsky2017pebbling} introduced comonads for pebble games;
\item this approach was further developed and extended to \EF~and bisimulation games in \cite{DBLP:conf/csl/AbramskyS18}. We recommend the extended version~\cite{AS2021}. 
\item More game comonads, including those for guarded and hybrid fragments, and generalised quantifiers, were defined in \cite{Guarded2021,Hybrid2022,conghaile2021game}.
\end{itemize}

The first wave culminated in the axiomatic framework of arboreal categories~\cite{AR2021icalp}; a more complete account can be found in the extended version~\cite{AR2022}. In the second wave, we see an emerging landscape in which structural features and dividing lines begin to appear, leading to:
\begin{itemize}
\item homomorphism counting results \emph{\`a la} Lov\'asz using game comonads~\cite{DJR2021}; the general pattern was extended beyond finite structures in~\cite{Reggio2022}. 
\item A uniform approach to homomorphism preservation theorems in logic based on arboreal categories~\cite{AR2024}, following the comonadic analysis of the Equirank-variable homomorphism preservation theorem in~\cite{Paine2020}.
\item An axiomatic study of Feferman--Vaught type theorems and Courcelle's
theorem~\cite{FVM2022,JMS2023}, and of graph parameters via density comonads~\cite{abramskyjaklpaine2022}.
\item Refinements of the axiomatic setting in the form of linear arboreal categories~\cite{AMS2024}, and the study of the expressive power of arboreal categories~\cite{reggio2023finitely}.
\item The use of sheaf-theoretic and cohomological methods in constraint satisfaction problems~\cite{abramsky2022notes,oconghaile:LIPIcs.MFCS.2022.75}, and the axiomatic study of further fragments of resource-sensitive logics~\cite{ALR23}.
\end{itemize}

In this paper we will outline the basic ideas underlying the theory of game comonads, and touch upon some of the more advanced topics, with the aim of giving the reader a glimpse of the rich landscape described above.

\subsection{Outline of the paper}
The paper is structured as follows. In~\S\S\ref{s:resources}--\ref{s:games} we introduce our running examples of logical resources and model-comparison games, and in~\S\ref{s:game-comonads} we define the corresponding game comonads. The first examples of logical equivalences that can be recovered from game comonads, and in particular those corresponding to the existential versions of the games, are illustrated in~\S\ref{s:logical-equiv}. In~\S\ref{s:coalgebras} we explain how coalgebras for game comonads encode combinatorial parameters of structures, and show how this can be used to obtain a uniform approach to \emph{homomorphism counting results} in finite model theory. 
The main focus of~\S\ref{s:open} is on Duplicator winning strategies in back-and-forth games, which can be encoded via an appropriate notion of \emph{open map bisimulation} between coalgebras. Finally, in~\S\S\ref{s:arboreal}--\ref{s:HPT} we give an overview of arboreal categories and their application to homomorphism preservation theorems in logic, and in~\S\ref{s:homotopical} we discuss some emerging links between resource-sensitive model theory and abstract homotopy.

\section{Deception and logical resources}
\label{s:resources}

In logic, models are typically studied not as they really are, i.e.\ up to isomorphism, but only up to definable properties; e.g., two models are elementarily equivalent if they satisfy the same first-order sentences. This idea is pervasive in mathematics and computer science: we may not want to distinguish between two topological spaces that are homotopy equivalent, or two programs that on the same input give the same output. 
From a logical viewpoint, every fragment $\LL\subseteq \FO$ of first-order logic induces an equivalence relation $\equiv_{\LL}$ on structures given by
\[
A\equiv_{\LL} B \ \ \Longleftrightarrow \ \ \text{for all } \phi\in\LL \ (A\models \phi \ \Leftrightarrow \ B\models\phi).
\]
The relation $\equiv_{\LL}$ allows us to compare structures up to $\LL$-definable properties.

When $\LL=\FO$, the relation $\equiv_{\LL}$ coincides with elementary equivalence. As any two elementarily equivalent finite structures are isomorphic, the main focus in finite model theory is on relations $\equiv_{\LL}$ that are strictly coarser than elementary equivalence. This can be achieved, for instance, by considering resource-bounded fragments of first-order logic. Intuitively, given a resource parameter $k\in\N$, we will define a corresponding stratification of $\FO$, i.e.\ an increasing chain of logic fragments 
\[
\LL_{0}\subseteq \LL_{1} \subseteq \LL_{2} \subseteq \cdots \subseteq \FO
\]
whose union is all of $\FO$. Typically, the equivalence relations $\equiv_{\LL_{k}}$ are strictly coarser than~$\equiv_{\FO}$, i.e.\ than elementary equivalence, but we recover the latter as
\[
\bigcap_{k\in\N}{\equiv_{\LL_{k}}} \ = \ \ {\equiv_{\FO}}.
\]

We illustrate two such stratifications of first-order logic, in which the resource parameters are given by the number of variables and the quantifier rank, respectively. We then outline a similar stratification in the case of modal logic. Throughout, we fix a countably infinite set of (pairwise distinct) first-order variables
\[
x_{1},x_{2},x_{3},\dots
\]
from which all first-order formulas are built. Moreover, for simplicity, we shall restrict ourselves to the case of finite relational signatures, i.e.\ signatures $\sigma$ containing finitely many relation symbols and constant symbols, but no function symbols. 

\subsection{Finite-variable logic} For each $n\in\N$, denote by 
\[
\FO^{n}
\] 
the set of all first-order sentences that only use the variables $x_{1},\ldots,x_{n}$. In particular, $\FO^{0}$ consists of the atomic sentences (such as $c_{1}=c_{2}$, $S(c_{1},c_{2},c_{3})$ and Boolean combinations thereof, where $c_{1},c_{2},c_{3}$ are constant symbols and $S$ is a relation symbol). Despite the severe syntactic restriction, the finite-variable fragments $\FO^{n}$ are remarkably expressive. This expressiveness comes from the fact that variables can be \emph{reused}. For example, for each integer $\ell\geq 1$ there is a three-variable sentence $\phi_{\ell}\in \FO^{3}$ in the language of graphs expressing the existence of a path of length $\ell$. Such a~$\phi_{\ell}$ can be defined as $\exists x_{1}, x_{2}. \psi_{\ell}$, where the formula $\psi_{\ell}(x_{1},x_{2})$ expresses the existence of a path of length $\ell$ from $x_{1}$ to $x_{2}$ and is defined as follows: 
\[
\psi_{1}(x_{1},x_{2}) \coloneqq R(x_{1},x_{2})
\]
and, for all $\ell>1$,
\[
\psi_{\ell}(x_{1},x_{2}) \coloneqq \exists x_{3}. (R(x_{1},x_{3}) \wedge \exists x_{1}. (x_{1} = x_{3} \wedge \psi_{\ell-1}(x_{1},x_{2}))).
\]

\subsection{Bounded quantifier rank} The \emph{quantifier rank} of a formula is the maximum number of nested quantifiers in the formula. For example, the formulas
\[
R(x_{1},x_{2}), \ \ \exists x_{2}. R(x_{1},x_{2}) \wedge \exists x_{3}. R(x_{3},x_{1})
\]
have quantifier rank $0$ and $1$, respectively, whereas the following has quantifier rank~$2$:
\[
\forall x_{2} \exists x_{3}. (R(x_{1},x_{3}) \wedge (R(x_{1},x_{2}) \to R(x_{3},x_{2}))).
\]
For each $k\in\N$, write 
\[
\FO_{k}
\] 
for the collection of all first-order sentences of quantifier rank at most~$k$. The following observation about the logic fragments $\FO_{k}$ will be important in the following; see e.g.\ \cite[Lemma~3.13]{Libkin2004} for a proof. (Recall that we are considering only \emph{finite} relational signatures.)
\begin{lemma}\label{l:qrank-finite}
For each $k\in\N$ there are, up to logical equivalence, finitely many formulas of quantifier rank at most $k$.
\end{lemma}
In particular, the quotient of $\FO_{k}$ with respect to logical equivalence is finite.
The analogous statement for the $n$-variable fragments $\FO^{n}$ is false in general: for example, for any two distinct integers $\ell,m\geq 1$, the sentences $\phi_{\ell},\phi_{m}\in \FO^{3}$ stating the existence of a path of length $\ell$ and $m$, respectively, are not equivalent.

\begin{remark}
We note in passing that logically equivalent formulas may have different quantifier rank or use a different number of variables. For example, the sentence
\[
\exists x_{1},\dots,x_{4}. (R(x_{1},x_{2}) \wedge R(x_{2},x_{3}) \wedge R(x_{3},x_{4}))
\]
uses $4$ variables and has quantifier rank $4$; it expresses the existence of a path of length~$3$, so it is equivalent to the sentence $\phi_{3} = \exists x_{1}, x_{2}. \psi_{3}$, which uses $3$ variables and has quantifier rank $6$.
\end{remark}

\subsection{Modal logic and modal depth} We consider modal logic as a fragment of first-order logic (in one free variable) via its standard translation. In more detail, let $\sigma$ be a \emph{modal signature}, i.e.\ a finite signature consisting of relation symbols of arity either $1$ or $2$. Each unary relation symbol ${P\in\sigma}$ yields a propositional variable~$p$, and each binary relation symbol $R_{\alpha}\in\sigma$ yields modalities $\Box_{\alpha}$ and $\Diamond_{\alpha}$.
Then a $\sigma$-structure $A$ corresponds to a Kripke structure for this \mbox{(multi-)}modal logic: the interpretation $P^A\subseteq A$ gives the valuation for the propositional variable $p$, and $R_{\alpha}^A\subseteq A^{2}$ gives the accessibility relation for $\Box_{\alpha}$ and $\Diamond_{\alpha}$.

Each formula $\phi$ in this modal logic admits a translation into a first-order formula $\tr{\phi}_x$ in one free variable $x$; this is known as the \emph{standard translation}, see e.g.~\cite[\S 2.4]{blackburn2002modal}. We let $\tr{p}_x\coloneqq P(x)$ and let $\tr{- }_x$ commute with Boolean connectives. Further, set 
\[
\tr{\Box_{\alpha}\phi}_x\coloneqq \forall y. \, R_{\alpha}(x,y) \to \tr{\phi}_y \ \ \text{ and } \ \ \tr{\Diamond_{\alpha}\phi}_x\coloneqq \exists y. \, R_{\alpha}(x,y) \wedge \tr{\phi}_y
\] where $y$ is a fresh variable. Modal formulas are evaluated in \emph{pointed Kripke structures}, that is pairs $(A,a)$ where $A$ is a $\sigma$-structure and $a\in A$. Then $A,a\models \phi$ according to Kripke semantics if, and only if, $A\models \tr{\phi}_x [a/x]$ in the standard model-theoretic sense.

In analogy with the quantifier rank of a first-order formula, we define the \emph{modal depth} of a modal formula as the maximum nesting of modalities in the formula. For example, $\neg p$, $\Diamond_{\alpha}p$ and $\Box_{\alpha}\Diamond_{\beta}p$ have modal depth $0$, $1$ and $2$, respectively. For each $k\in\N$, we denote by 
\[
\ML_{k}
\] 
the collection of all modal formulas with modal depth at most $k$. The modal depth of a modal formula coincides with the quantifier rank of its standard translation, hence Lemma~\ref{l:qrank-finite} implies that each modal fragment $\ML_{k}$ contains only finitely many formulas, up to logical equivalence.

\section{Model-comparison games}
\label{s:games}

In this section we will introduce three examples of model-comparison games: the bisimulation game for modal logic, and the \EF~and pebble games for first-order logic. All are two-player games, played by Spoiler and Duplicator on $\sigma$-structures $A$ and $B$. While Spoiler aims to show that $A$ and $B$ are different, Duplicator aims to show that they are similar, i.e.\ that they satisfy the same definable properties. 

As before, we assume that $\sigma$ is a finite relational signature (or a modal signature, in the case of modal logic).

\subsection{Bisimulation games}
Fix $k\in\N$. The $k$-round bisimulation game for modal logic is played by Spoiler and Duplicator on pointed Kripke structures $(A,a)$ and $(B,b)$ as follows. The initial position is $(a_{0},b_{0})\coloneqq (a,b)$. At round $i+1$ (with $0\leq i < k$) with current position $(a_{i},b_{i})$, Spoiler chooses one of the two structures and a binary relation symbol in the signature, say $A$ and $R_{\alpha}$, and an element $a_{i+1}\in A$ such that $R^{A}_{\alpha} (a_{i}, a_{i+1})$. Duplicator must respond by choosing an element in the other structure, in this case $b_{i+1}\in B$, such that $R^{B}_{\alpha} (b_{i}, b_{i+1})$. If Duplicator has no such response available, they lose. Otherwise, Duplicator wins the $k$-round bisimulation game if, for all unary predicates~$P$ in the signature, we have
\[
\forall i\in \{0,\ldots,k\}, \ \ P^{A}(a_{i}) \Leftrightarrow P^{B}(b_{i}).
\] 

The following classical result links the existence of a Duplicator winning strategy in this game with equivalence in the bounded-depth fragment of modal logic:
\begin{theorem}[\cite{HM1980}]\label{t:game-logic-modal}
Duplicator has a winning strategy in the $k$-round bisimulation game played between $(A,a)$ and $(B,b)$ if, and only if, $(A,a)$ and $(B,b)$ satisfy the same formulas in $\ML_{k}$.
\end{theorem}

\subsection{\EF~games}
As before, fix $k\in\N$. The $k$-round \EF~game, played by Spoiler and Duplicator on structures $A$ and $B$, can be described as follows. In the $i$th round (with $1\leq i\leq k$), Spoiler chooses an element in one of the two structures, say $a_{i}\in A$, and Duplicator responds by choosing an element in the other structure, say $b_{i}\in B$. Duplicator wins after $k$ rounds if the relation \[\{(a_{i},b_{i}) \mid 1 \leq i \leq k\}\] is a partial isomorphism, i.e.\ the map $a_{i}\mapsto b_{i}$ yields an isomorphism between the induced substructures of $A$ and $B$ determined by the sets $\{a_{1},\ldots,a_{k}\}$ and $\{b_{1},\ldots,b_{k}\}$, respectively. In particular, note that Duplicator wins the $0$-round game precisely when $A$ and $B$ satisfy the same atomic sentences. Furthermore, for any $j\in \N$, Duplicator wins the $(j+1)$-round game if, and only if, the following conditions are satisfied:
\begin{enumerate}[label=(\roman*)]
\item\label{i:forth-cond} For every $a\in A$ there exists $b\in B$ such that Duplicator wins the $j$-round game between the pointed structures $(A,a)$ and $(B,b)$;\footnote{In other words, we expand the signature by adding a fresh constant symbol $\mathfrak{c}$, and regard $A$ and $B$ as structures for this new signature by setting $\mathfrak{c}^{A}\coloneqq a$ and $\mathfrak{c}^{B}\coloneqq b$. The $j$-round game is then played on the extended structures.}
\item\label{i:back-cond} For every $b\in B$ there exists $a\in A$ such that Duplicator wins the $j$-round game between the pointed structures $(A,a)$ and $(B,b)$.
\end{enumerate}

The existence of a winning strategy for Duplicator captures precisely equivalence in first-order logic with bounded quantifier rank:
\begin{theorem}[\cite{Ehrenfeucht1960,Fraisse1954}]\label{th:EF}
Duplicator has a winning strategy in the $k$-round \EF~game played between $A$ and $B$ if, and only if, $A$ and $B$ satisfy the same first-order sentences in $\FO_{k}$.
\end{theorem}

We briefly sketch a proof of Theorem~\ref{th:EF}, which proceeds by induction on $k$. The base case $k=0$ is obvious because the first-order sentences with quantifier rank $0$ are exactly the atomic ones. Next, we show that if Theorem~\ref{th:EF} holds for $k$, it also holds for $k+1$. To this end, for all elements $a$ of a structure $A$, denote by $\tp_{k}(a)$ the \emph{rank-$k$ type of $a$}, i.e.\ the set of all first-order formulas $\phi(x)$ with quantifier rank $\leq k$ such that $A\models \phi(a)$. 

In one direction, suppose $A$ and $B$ satisfy the same first-order sentences with quantifier rank $\leq k+1$. We must show that Duplicator has a winning strategy in the $(k+1)$-round game between $A$ and $B$. We will check condition~\ref{i:forth-cond} above; the proof of~\ref{i:back-cond} is similar. Pick an arbitrary element $a\in A$ and consider its rank-$k$ type $\tp_{k}(a)$. We have
\[
A\models \bigwedge{\tp_{k}(a)}.
\]
Although the latter conjunction is infinite, it follows from Lemma~\ref{l:qrank-finite} that it is equivalent to a single formula $\phi_{a}(x)$ of quantifier rank $\leq k$. Hence
\[
A\models \exists x. \phi_{a}(x),
\]
and since the sentence $\exists x. \phi_{a}(x)$ has quantifier rank $\leq k+1$ we get
\[
B\models \exists x. \phi_{a}(x).
\]
If $b\in B$ is a witness for the latter existential quantifier, we have $\tp_{k}(a) = \tp_{k}(b)$. Thus $(A,a)$ and $(B,b)$ satisfy the same first-order sentences with quantifier rank $\leq k$ and so, by inductive hypothesis, Duplicator wins the $k$-round game between $(A,a)$ and $(B,b)$.

The other direction follows by observing that every sentence in $\FO_{k+1}$ is a Boolean combination of sentences of the form $\exists x. \phi(x)$ where $\phi(x)$ has quantifier rank $\leq k$, and using items~\ref{i:forth-cond}--\ref{i:back-cond} above. Cf.\ e.g.\ \cite[Theorem~3.18]{Libkin2004}.

\subsection{Pebble games}\label{s:pebble-games}
Fix $n\in\N$. In the $n$-pebble game between structures $A$ and $B$, Spoiler and Duplicator each have $n$ pebbles available. In the $i$th round, Spoiler places some pebble $p_{i}$ on an element of one of the two structures, say $a_{i}\in A$, and Duplicator responds by placing their corresponding pebble $p_{i}$ on an element of the other structure, say $b_{i}\in B$. If the pebble $p_{i}$ was previously placed on some other element, the effect is to move it to the newly chosen element (this corresponds to the fact that variables in a formula can be reused). 
After $k$ rounds have been played, we have sequences 
\[
[(p_{1}, a_{1}),\ldots,(p_{k}, a_{k})] \ \ \text{ and } \ \ [(p_{1}, b_{1}),\ldots,(p_{k}, b_{k})]
\] 
that record the placings of pebbles on elements during the play. The current placing of pebble $p$ is the last element in the sequence with first component $p$.\footnote{The current placings of the pebbles can be thought of as ``windows'' of size bounded by $n$ onto the structures. These windows can slide around over different parts of the structures as moves are played.} Duplicator wins the round if the relation determined by the current placings of the pebbles is a partial isomorphism. Duplicator wins the $n$-pebble game if they have a strategy that is winning after $k$ rounds, for all $k\in\N$. 

Note that, whereas the resource parameter in the bisimulation and \EF~games is the number of rounds, pebble games are infinite and the relevant resource parameter is the number of pebbles.
The infinite nature of pebble games entails that the existence of a winning strategy for Duplicator captures equivalence in an \emph{infinitary extension} of first-order logic: 

\begin{definition}
Let $L_{\infty,\omega}$ be the set of all sentences using, in addition to the usual first-order connectives, conjunctions and disjunctions of any cardinality.\footnote{The two subscripts in $L_{\infty,\omega}$ indicate that we allow infinitary conjunctions and disjunctions ($\infty$) but only finitary quantifications ($\omega$). Note that for all sentences $\phi\in L_{\infty,\omega}$, each subformula of $\phi$ contains only finitely many variables.} We write $L^{n}_{\infty,\omega}$ for the fragment of $L_{\infty,\omega}$ consisting of those sentences in the variables $x_{1},\ldots,x_{n}$.
\end{definition}

\begin{theorem}[\cite{Barwise1977,Immerman1982}]\label{th:pebble} 
Duplicator has a winning strategy in the $n$-pebble game played between $A$ and $B$ if, and only if, $A$ and $B$ satisfy the same sentences in $L^{n}_{\infty,\omega}$.
\end{theorem}

If the structures $A$ and $B$ are finite, then every infinite strategy in the $n$-pebble game is determined by a $k$-round strategy for a sufficiently large $k$, namely the one in which all finitely many possible placings of the pebbles appeared. Therefore, the equivalence relations $\equiv_{L^{n}_{\infty,\omega}}$ and $\equiv_{\FO^{n}}$ coincide over finite structures; cf.\ e.g.\ \cite[Corollary~2.19]{KV1992} or \cite[Proposition~11.8]{Libkin2004}.

\begin{corollary}\label{cor:pebble-fin-struct}
Let $A$, $B$ be finite structures. Duplicator has a winning strategy in the $n$-pebble game played between $A$ and $B$ if, and only if, $A$ and $B$ satisfy the same sentences in $\FO^{n}$.
\end{corollary}

\section{A new perspective: Game comonads}
\label{s:game-comonads}

Game comonads arise from the insight that model-comparison games can be viewed as semantic constructions in their own right. More precisely, given a notion of game $G$ and a resource parameter~$k$, for each structure $A$ we construct a structure $\G_{k}A$ (in the same signature) whose universe consists of all possible plays in $A$ in the game $G$ with $k$ resources. What is more, the assignment $A\mapsto \G_{k} A$ is functorial and yields a \emph{comonad} on the appropriate category of structures. 

We start by recalling the notion of comonad, which is dual to that of a monad:

\begin{definition}
A \emph{comonad (in Kleisli form)} on a category $\C$ is given by:
\begin{itemize}
\item an object map $\G\colon \Ob(\C)\to \Ob(\C)$,
\item a morphism $\epsilon_A\colon \G A\to A$ for every object $A$,
\item and a \emph{coextension operation} associating with each morphism $f\colon \G A\to B$ a morphism $f^*\colon \G A\to \G B$,
\end{itemize}
such that for all morphisms $f\colon \G A\to B$ and $g\colon \G B\to C$, the following equations hold:
\[
\epsilon_A^*=\id_{\G A}, \ \ \ \ \ \epsilon_B\circ f^*=f, \ \ \ \ \ (g\circ f^*)^*=g^*\circ f^*.
\]
\end{definition}

We can always extend a comonad in Kleisli form $\G$ to a functor $\C\to\C$ by setting $\G f \coloneqq(f\circ \epsilon_A)^*$ for every morphism $f\colon A\to B$. Moreover, setting $\delta_A\coloneqq \id_{\G A}^*$ for every $A$ in $\C$, the tuple $(\G,\epsilon,\delta)$ is a comonad in the more traditional sense, where the natural transformations $\epsilon$ and $\delta$ are the \emph{counit} and the \emph{comultiplication} of the comonad, respectively. In fact, these two formulations are equivalent.

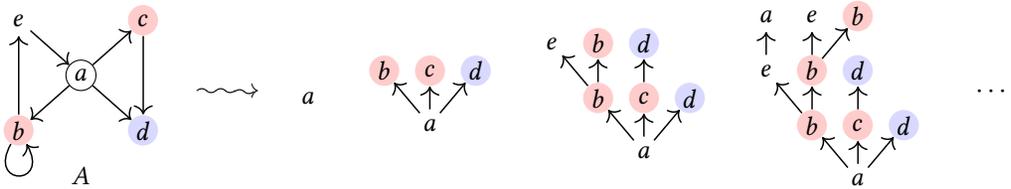
\begin{figure}[h]
\[
\begin{tikzcd}[column sep=small,row sep=small,cells = {nodes={circle,
                minimum size=4mm, inner sep=-1pt, anchor=center}}]
	e && |[fill=red!20]| c \\
	& |[draw]| a \\
	|[fill=red!20]| b \arrow[out=240,in=300,loop] && |[fill=blue!15]| d \\ [-3.5pt] 
	&A&
	\arrow[from=2-2, to=1-3]
	\arrow[from=2-2, to=3-3]
	\arrow[from=1-3, to=3-3]
	\arrow[from=2-2, to=3-1]
	\arrow[from=3-1, to=1-1]
	\arrow[from=1-1, to=2-2]
\end{tikzcd}
\ \ \ \longsquiggly \ \ \
\begin{tikzcd}[column sep=tiny,row sep=small,cells = {nodes={circle,
                minimum size=4mm, inner sep=-1pt, anchor=center}}]
	a
\end{tikzcd}
\ \ \ \ \ \ \ 
%
\begin{tikzcd}[column sep=tiny,row sep=small,cells = {nodes={circle,
                minimum size=4mm, inner sep=-1pt, anchor=center}}]
	|[fill=red!20]| b & |[fill=red!20]| c & |[fill=blue!15]| d \\
	& a
	\arrow[from=2-2, to=1-2]
	\arrow[from=2-2, to=1-3]
	\arrow[from=2-2, to=1-1]
\end{tikzcd}
\ \ \ \ \ \ \ 
%
\begin{tikzcd}[column sep=tiny,row sep=small,cells = {nodes={circle,
                minimum size=4mm, inner sep=-1pt, anchor=center}}]
	e & |[fill=red!20]| b & |[fill=blue!15]| d \\
	& |[fill=red!20]| b & |[fill=red!20]| c & |[fill=blue!15]| d \\
	&& a
	\arrow[from=3-3, to=2-4]
	\arrow[from=3-3, to=2-3]
	\arrow[from=3-3, to=2-2]
	\arrow[from=2-3, to=1-3]
	\arrow[from=2-2, to=1-2]
	\arrow[from=2-2, to=1-1]
\end{tikzcd}
\ \ \ \ \ \ \ 
\begin{tikzcd}[column sep=tiny,row sep=small,cells = {nodes={circle,
                minimum size=4mm, inner sep=-1pt, anchor=center}}]
	a & e & |[fill=red!20]| b \\
	e & |[fill=red!20]| b & |[fill=blue!15]| d \\
	& |[fill=red!20]| b & |[fill=red!20]| c & |[fill=blue!15]| d \\
	&& a
	\arrow[from=4-3, to=3-4]
	\arrow[from=4-3, to=3-3]
	\arrow[from=4-3, to=3-2]
	\arrow[from=3-3, to=2-3]
	\arrow[from=3-2, to=2-2]
	\arrow[from=3-2, to=2-1]
	\arrow[from=2-2, to=1-3]
	\arrow[from=2-2, to=1-2]
	\arrow[from=2-1, to=1-1]
\end{tikzcd}
\qquad \cdots
\]
\centering
\caption{A pointed Kripke model $(A,a)$ and the first terms of its sequence of $k$-unravellings (for $k=0,1,2,3$).}
\label{fig:unravelling}
\end{figure}

\vspace{1em}
The structure $\G_{k}A$ can be thought of as a forest-ordered \emph{cover}, or \emph{decomposition}, of~$A$. A fundamental example of this idea arises from the unravelling construction in modal logic, which we now recall.  
Consider the pointed Kripke model $(A,s)$ in Figure~\ref{fig:unravelling}; the modal signature consists of a single binary relation symbol (represented by an arrow) and two unary relation symbols (represented as red and blue nodes).
The \emph{unravelling} of $(A,a)$ is the Kripke model of all possible finite paths in $A$ starting from~$x$; see \S\ref{s:modal-comonad} for a precise definition. For each $k\in\N$, the \emph{$k$-unravelling} of $(A,a)$ is obtained by restricting to paths of length at most~$k$. Note that the universe of the $k$-unravelling of $(A,a)$ can be identified with the set of all possible plays in $(A,a)$ in the $k$-round bisimulation game.

After introducing the relevant categories of relational structures on which game comonads are defined, we will show that the unravelling construction yields a comonad on pointed Kripke models, the so-called \emph{modal comonad}, and introduce two further examples of game comonads: the \emph{\EF}~and \emph{pebbling comonads}. These comonads can be regarded as enrichments of the \emph{finite-list comonad} on the category of sets, and variations thereof.

\subsection{Categories of relational structures}
Henceforth, we shall fix a finite relational signature $\sigma$; we will assume for simplicity that $\sigma$ contains no constant symbols.\footnote{Note, however, that this assumption prevents us from encoding the distinguished element of a pointed Kripke model by means of a constant symbol in the signature.} Hence, a $\sigma$-structure consists of a set $A$ together with, for each relation symbol $R\in \sigma$ of arity $n$ (with $n$ an integer greater than $0$), a subset $R^{A}\subseteq A^{n}$. A \emph{homomorphism} from a $\sigma$-structure $A$ to a $\sigma$-structure $B$ is given by a function $h\colon A\to B$ such that, for each $R\in\sigma$ of arity $n$ and $(a_{1},\ldots,a_{n})\in A^{n}$,
\[
(a_{1},\ldots,a_{n})\in R^{A} \ \Longrightarrow \ (h(a_{1}),\ldots,h(a_{n}))\in R^{B}.
\]

Our setting will be the category \[\CS\] of $\sigma$-structures and their homomorphisms. Note that this is \emph{not} the usual setting for classical model theory, which is mainly concerned with morphisms that preserve and reflect the validity of first-order sentences, such as elementary embeddings. Nevertheless, homomorphisms of relational structures are pervasive in computer science, e.g.\ in databases, constraint satisfaction and finite model theory, and also in combinatorics (in the study of graphs and graph homomorphisms). 

When dealing with modal logic, we will assume that $\sigma$ is a modal signature and consider the category \[\CSstar\] of pointed Kripke models whose morphisms are the homomorphisms that preserve the distinguished elements.

\subsection{The modal comonad}\label{s:modal-comonad}

For each pointed Kripke model $(A,a)\in\CSstar$, we define a new model $\Mo(A,a)\in\CSstar$ as follows. The underlying set of $\Mo(A,a)$ consists of all finite paths in $A$ starting from $a$:
\[
a \xrightarrow{R_{1}} a_{1} \xrightarrow{R_{2}} a_{2} \xrightarrow{\phantom{R_{3}}} \cdots \xrightarrow{R_{\ell}} a_{\ell}
\]
where $R_{1},\ldots,R_{\ell}$ are binary relation symbols in $\sigma$. These finite paths can be thought of as plays in the bisimulation game. The distinguished element of $\Mo(A,a)$ is the trivial path $[a]$ of length $0$, and the interpretation in $\Mo(A,a)$ of the relation symbols is defined as follows:
\begin{itemize}
\item If $P\in\sigma$ is a unary relation symbol, its interpretation consists of the paths whose last element is in $P^{A}$.
\item If $R\in\sigma$ is a binary relation symbol, its interpretations consists of the pairs of paths $(p,p')$ such that $p'$ is obtained by extending $p$ by one step along $R$.
\end{itemize}
The pointed Kripke model $\Mo(A,a)$ is the \emph{(full) unravelling of} $(A,a)$. 

The assignment $(A,a)\mapsto \Mo(A,a)$ gives an object map from $\CSstar$ to itself, and moreover we have homomorphisms of pointed Kripke models
\[
\epsilon_{(A,a)}\colon \Mo(A,a)\to (A,a)
\]
sending a path to its last element. The coextension operation associating with a morphism $f\colon \Mo(A,a)\to (B,b)$ in $\CSstar$ a morphism
\[
f^{*}\colon \Mo(A,a)\to \Mo(B,b)
\]
is defined recursively: $f^{*}$ sends the trivial path $(a)$ to $(b)$, and if $p'\in \Mo(A,a)$ is obtained by extending $p$ by one step along $R$ to $a'\in A$, then $f^{*}(p')$ is the sequence that extends $f^{*}(p)$ by one step along $R$ to $f(p')$.
These data yield a comonad on the category $\CSstar$ of pointed Kripke models: 

\begin{proposition}
$\Mo$ is a comonad on $\CSstar$.
\end{proposition} 

Intuitively, the comonad $\Mo$ corresponds to the \emph{infinite} bisimulation game. It is often useful to consider resource-bounded variants of this construction, where the length of paths is bounded; this corresponds to considering finite bisimulation games. Accordingly, for each $k\in\N$, we let $\Mk(A,a)$ be the submodel of $\Mo(A,a)$ defined by the paths of length at most $k$---this is known as the \emph{$k$-unravelling} of $(A,a)$. The morphisms $\epsilon_{(A,a)}$ and the coextension operation above restrict to $\Mk(A,a)$ in the obvious manner. In fact,
\begin{proposition}
$\Mk$ is a comonad on $\CSstar$, for each $k\in\N$.
\end{proposition}

We refer to the comonads $\Mo$ and $\Mk$ as \emph{modal comonads}. 
Modal comonads have an important property that is not shared by all game comonads: they are \emph{idempotent}, meaning that the components of the comultiplication $\delta\colon \Mo\Rightarrow\Mo\Mo$ are isomorphisms. Informally, this can be understood as the fact that the unravelling construction gives a universal way of constructing a tree-like Kripke model from an arbitrary one, and applying this construction to a model that is already tree-like yields nothing new. In particular, $\Mo(A,a)$ is isomorphic to $\Mo\Mo(A,a)$. 

\begin{remark}
Note that the unravelling $\Mo(A,a)$ can be infinite even if $A$ is finite; e.g., the unravelling of a model with one node and one loop is a countably infinite chain. On the other hand, for each $k\in\N$, the $k$-unravelling $\Mk(A,a)$ is finite whenever $A$ is.

In the other direction, it is possible that $A$ is infinite but $\Mo(A)$ is finite; e.g., take~$A$ to be the disjoint union of a node $a$ (the distinguished element) and an infinite chain not reachable from $a$. In fact, the unravelling construction ignores all parts of a model that are not reachable from its distinguished element.
\end{remark}

\subsection{The \EF\ comonad}
For each relational structure $A\in\CS$, we define a new structure $\Eo(A)$ whose underlying set consists of all non-empty finite sequences of elements from $A$. This choice is motivated by the fact that, while in the bisimulation game the players can only move along accessibility relations, in the \EF~game they can move freely within the structures. 

As before, we denote by 
\[
\epsilon_{A}\colon \Eo(A)\to A, \ \ \epsilon_{A}[a_{1},\ldots,a_{\ell}]\coloneqq a_{\ell}
\]
the map sending a sequence to its last element. If $R\in\sigma$ is a relation symbol of arity $n$, then its interpretation in $\Eo(A)$ consists of the tuples $(s_{1},\ldots,s_{n})$ of sequences verifying the following two conditions:

\begin{itemize}
\item the sequences $s_{1},\ldots,s_{n}$ are pairwise comparable in the prefix order (i.e., for all $i,j\in \{1,\ldots,n\}$, either $s_{i}$ is a prefix of $s_{j}$ or vice versa), and
\item $(\epsilon_{A}(s_{1}),\ldots,\epsilon_{A}(s_{n}))\in R^{A}$, i.e.\ the last elements of the sequences are $R$-related in $A$.
\end{itemize}
It then follows that the $\epsilon_{A}$ are homomorphisms. The coextension operation sends
a homomorphism $f \colon \Eo(A) \to B$ to the homomorphism
\[
f^{*}\colon \Eo(A) \to \Eo(B), \ \ f^{*}[a_{1},\ldots, a_{\ell}] \coloneqq [b_{1},\ldots, b_{\ell}]
\]
where $b_{i} \coloneqq f[a_{1},\ldots,a_{i}]$ for all $1 \leq i \leq \ell$.

\begin{proposition}\label{p:E-omega}
$\Eo$ is a comonad on $\CS$.
\end{proposition} 
In a similar way to the modal case, the comonad $\Eo$ corresponds to the infinite \EF~game. To capture finite games, for each $k\in\N$ we consider the submodel $\Ek(A)$ of $\Eo(A)$ consisting of sequences of length at most $k$. With the obvious restrictions of the morphisms $\epsilon_{A}$ and of the coextension operation, we have that
\begin{proposition}
$\Ek$ is a comonad on $\CS$, for each $k\in\N$.
\end{proposition} 

We refer to the comonads $\Eo$ and $\Ek$ as \emph{\EF~comonads}. 

\begin{remark}
The structure $\Eo(A)$ is always infinite whereas, for each $k\in \N$, $\Ek(A)$ is finite if, and only if $A$ is finite. The ``only if'' part of the latter statement follows from the fact that the components of the counit 
\[
\epsilon_{A}\colon \Ek(A)\to A
\]
are surjective. This feature distinguishes the \EF~comonads from the modal comonads, for which the components of the counit are generally not surjective because their images cannot contain unreachable parts of the model (or even parts of the model only reachable in more than $k$ steps, in the case of the comonads $\Mk$).
\end{remark}

\subsection{The pebbling comonad}
To encode pebble games, we follow the same pattern we adopted for bisimulation and \EF~games. For each $n\in\N$, let $\n\coloneqq\{1,\ldots,n\}$. For each relational structure $A\in\CS$, we define a structure $\Pn(A)\in\CS$ whose underlying set consists of all non-empty finite sequences of elements from the Cartesian product of sets $\n\times A$. An element $(p,a)\in \n\times A$ can be thought of as the move in which pebble $p$ is placed on $a$, and a finite sequence of such moves corresponds to a play in the $n$-pebble game. We refer to $p$ as the \emph{pebble index} of the move $(p,a)$. 

As before, we have a map
\[
\epsilon_{A}\colon \Pn(A) \to A, \ \ \epsilon_{A}[(p_{1},a_{1}), \ldots, (p_{\ell},a_{\ell})]\coloneqq a_{\ell}
\]
sending a play to the element selected in the last move. For each relation symbol $R\in\sigma$ of arity~$j$, its interpretation in $\Pn(A)$ consists of the tuples of moves $(s_{1},\ldots,s_{j})$ satisfying the following properties:

\begin{itemize}
\item the sequences $s_{1},\ldots,s_{j}$ are pairwise comparable in the prefix order,
\item for all $i,i'\in \{1,\ldots, j\}$, whenever $s_{i}$ is a prefix of $s_{i'}$, the pebble index of the last move in $s_{i}$ does not appear in the suffix of $s_{i}$ in $s_{i'}$,\footnote{That is, the pebble has not been moved and is therefore still ``active''.} and 
\item $(\epsilon_{A}(s_{1}),\ldots,\epsilon_{A}(s_{j}))\in R^{A}$.
\end{itemize}
With this definition, the functions $\epsilon_{A}$ are homomorphisms. The coextension operation is the same, mutatis mutandis, as with $\Eo$. We have:

\begin{proposition}
$\Pn$ is a comonad on $\CS$, for each $n\in\N$.
\end{proposition} 

We refer to the comonads $\Pn$ as \emph{pebbling comonads}. 

\begin{remark}
Since there is no bound on the length of plays, $\Pn(A)$ is always infinite, even when $A$ is finite. This cannot be avoided: there is no comonad that captures pebble games and sends finite structures to finite structures (for a formal statement, see \cite[Theorem~7]{abramsky2017pebbling}). 

On the other hand, we can stratify $\Pn(A)$ by considering the submodels defined by sequences of length at most $k$, for each $k\in \N$; this corresponds to playing $k$-round $n$-pebble games. The ensuing comonads $\mathbb{P}_{n,k}$ on $\CS$ do restrict to finite structures.
\end{remark}

\section{The Kleisli category and the first logical equivalences}
\label{s:logical-equiv}

Game comonads provide a categorical view on a number of model-comparison games, which in turn capture in a combinatorial fashion preservation of corresponding logic fragments. In this section and in \S\ref{s:open} below, we will see how preservation of many logic fragments can be detected directly at the categorical level. In this section we look at the \emph{Kleisli categories} for game comonads, and show that two natural relations in these categories---the homomorphism preorder and the isomorphism relation---capture preservation of \emph{existential positive} fragments and extensions with \emph{counting quantifiers}, respectively. These results hold uniformly for the game comonads we have introduced (\EF, pebbling and modal comonads) and the corresponding logical resources (quantifier rank, number of variables and modal depth).

In \S\ref{s:open} we will see that other logic fragments, including the full fragments $\FO_{k}$, $\FO^{n}$ and $\ML_{k}$, can be captured categorically by working in an enlargement of the Kleisli categories, namely the categories of coalgebras for the comonads. 

\subsection{The Kleisli category}

\begin{definition}
Let $\G$ be a comonad on a category $\C$. The \emph{Kleisli category of $\G$}, denoted by $\KL(\G)$, is defined as follows:

\begin{itemize}
\item $\Ob(\KL(\G)) \coloneqq \Ob(\C)$.
\item An arrow $A\to B$ in $\KL(\G)$, called a \emph{Kleisli arrow}, is an arrow $\G A\to B$ in $\C$. 
\end{itemize}
The Kleisli identity $A\to A$ is the arrow $\epsilon_{A}\colon \G A\to A$ in $\C$, and the composite of Kleisli arrows $A\to B\to C$ is defined using the coextension operation: given $f\colon \G A\to B$ and $g\colon \G B\to C$ in $\C$, we get $g\circ f^{*}\colon \G A \to \G B \to C$.
\end{definition}

The Kleisli category $\KL(\G)$ effectively considers the same objects as $\C$ but with a different notion of morphism. For example, an arrow $A\to B$ in the Kleisli category $\KL(\Ek)$ for the \EF~comonad is a homomorphism of $\sg$-structures  
\[
\Ek(A) \to B.
\]
The structure $\Ek(A)$ consists of ``explorations of depth at most $k$'' of~$A$, and reflects the finite amount of resources available to the players (in this case, the number of rounds).
Hence, the passage from the category $\CS$ to the Kleisli category for a game comonad (and, more generally, to the category of coalgebras) allows us to work in a setting that is intrinsically resource-sensitive.
For instance, we shall see in \S\ref{s:epf} that Kleisli arrows encode preservation of resource-bounded existential positive fragments. Before that, we briefly explain how to deal with the equality symbol.

\subsection{The equality symbol}\label{s:equality}
In their basic form, game comonads allow us to capture fragments of logics \emph{without} equality. This is sufficient for modal logic, since the image of the standard translation is contained in the equality-free fragment of $\FO$. On the other hand, to model logics \emph{with} equality, such as $\FO_{k}$ and $\FO^{n}$ (and variations thereof), we can proceed as follows.

Consider a (finite) relational signature $\sigma$, a fresh binary relation symbol $I$ and the expanded signature
\[
\sigma^{I}\coloneqq \sigma \cup \{I\}.
\]
There is a fully faithful functor $J\colon \CS \to \CSplus$ that upgrades a $\sg$-structure to a $\sg^{I}$-structure by interpreting $I$ as the diagonal relation. This functor has a left adjoint $H\colon \CSplus\to\CS$ that sends a $\sg^{I}$-structure $A$ to the quotient $\underline{A}\, / {\sim}$, where $\underline{A}$ is the $\sg$-reduct of $A$ and $\sim$ is the equivalence relation generated by $I^{A}$. Intuitively, the functor $J$ introduces (the interpretation of) the equality symbol, while $H$ eliminates it. Because a generic game comonad $\G$ is defined uniformly for any relational signature, we can consider its variant $\G^{I}$ over $\sg^{I}$-structures:
\[\begin{tikzcd}[column sep=0.8em]
\CS \arrow[hbend left]{rr}{J} & \text{\footnotesize{$\top$}} & \CSplus \arrow[hbend left]{ll}{H} \arrow[loop right, looseness=6, out=12, in=-12, "\G^{I}"]
\end{tikzcd}\]
The composite 
\[
\G^{=}\coloneqq H\circ \G^{I} \circ J \colon \CS \to \CS
\]
is also a comonad, and it is the one that will allow us to capture logics with equality.

Consider e.g.\ the \EF~comonad $\Ek$. Note that, for any $\sg$-structure~$A$, the interpretation of the relation $I$ in the $\sg^{I}$-structure $\EkI J(A)$ consists of those pairs $(s,t)$ of sequences of length at most $k$ of elements of $A$ such that (i) $s,t$ are comparable in the prefix order $\sqsubseteq$ and (ii) their last elements are equal. In other words, the relation $I^{\EkI J(A)}$ detects repeating sequences. Given $\sg$-structures~$A$ and~$B$, Kleisli morphisms 
\[
\Ek^{=}(A) \to B
\]
are in bijection with morphisms 
\[
\EkI J(A) \to J(B)
\]
(because $H\dashv J$), which in turn are in bijection with Kleisli morphisms $f\colon \Ek (A)\to B$ such that
\[
\text{$\forall s,t\in \Ek(A)$, \ if $s\sqsubseteq t$ and $\epsilon_{A}(s)=\epsilon_{A}(t)$ then $f(s)=f(t)$},
\]
called \emph{$I$-morphisms}. In the case of the pebbling comonad, $I$-morphisms are defined in a similar fashion.

\begin{remark}\label{rem:comparison-functor-Kleisli}
There is a fully faithful functor 
\[
\KL(\G^{=})\to \KL(\G^{I})
\]
that sends a $\sg$-structure $A$ to $J(A)$. So, given any two $\sg$-structures $A$ and $B$, there exists an (iso)morphism $A\to B$ in $\KL(\G^{=})$ if, and only if, there exists an (iso)morphism $J(A)\to J(B)$ in $\KL(\G^{I})$.
\end{remark}

The approach outlined above for encoding the equality symbol, whereby one ``composes'' a game comonad with the adjunction $H\dashv J$, is equivalent to the one based on $I$-morphisms introduced in \cite{abramsky2017pebbling,AS2021} and is a special case of the notion of \emph{relative adjunction} \cite{Ulmer1968}.

\subsection{Existential positive fragments}\label{s:epf}
An \emph{existential positive} formula is a formula that contains no negations and no universal quantifiers. In the case of modal formulas, this amounts to barring the use of the negation symbol and the modalities $\Box_{\alpha}$. Given a class of first-order formulas $\LL$ or, more generally, of formulas in infinitary first-order logic $L_{\infty,\omega}$, we denote by \[\exists^{+}\LL\] the subset of $\LL$ consisting of the existential positive formulas.

The next result states that the homomorphism preorder in the Kleisli categories of game comonads captures precisely preservation of the existential positive fragments. In this case, we do not need to consider the variant $\G^{=}$ of a game comonad $\G$ because the existential positive fragments admit \emph{equality elimination}; see Remark~\ref{rem:eq-elim-ep}.
Given a logic fragment $\LL$ and structures $A$ and $B$, we write 
\[
A\IMP_{\LL} B
\]
provided that $A\models\phi$ implies $B\models \phi$ for all $\phi\in\LL$. (Hence, the equivalence relation $\equiv_{\LL}$ is the symmetrization of the preorder $\IMP_{\LL}$.)
\begin{theorem}[\cite{abramsky2017pebbling,AS2021}]\label{th:ep-Kleisli-morphisms}
The following statements hold for each $k,n\in\N$ and (pointed) structures $A,B$: 
\begin{enumerate}
\item $A\IMP_{\EPFO_{k}} B$ if, and only if, there is an arrow $A\to B$ in $\KL(\Ek)$.
\item\label{i:ep-n-var} $A\IMP_{\exists^{+}L^{n}_{\infty,\omega}} B$ if, and only if, there is an arrow $A\to B$ in $\KL(\Pn)$.
\item $(A,a)\IMP_{\EPML_{k}} (B,b)$ if, and only if, there is an arrow $(A,a)\to (B,b)$ in $\KL(\Mk)$.
\end{enumerate}
\end{theorem}

\begin{remark}
If $A,B$ are finite, the relation $A\IMP_{\exists^{+}L^{n}_{\infty,\omega}} B$ in item~\ref{i:ep-n-var} above can be replaced with $A\IMP_{\EPFO^{n}} B$ (cf.\ \S\ref{s:pebble-games} and \cite[Corollary~4.9]{KV1995}).
\end{remark}

We briefly illustrate the idea behind Theorem~\ref{th:ep-Kleisli-morphisms} in the case of \EF~games; the same reasoning applies, mutatis mutandis, to the other cases. 

It is well know that the relation $A\IMP_{\EPFO_{k}} B$ holds precisely when Duplicator wins the \emph{existential positive} variant of the $k$-round \EF~game played between $A$ and $B$. In this game, Spoiler is forced to play in $A$ and Duplicator responds in $B$. The winning condition, after $k$ rounds, is that the relation \[\{(a_{i},b_{i}) \mid 1 \leq i \leq k\}\] determined by the chosen elements is a partial homomorphism (rather than a partial isomorphism), i.e.\ the map $a_{i}\mapsto b_{i}$ yields a homomorphism between the induced substructures of $A$ and $B$ determined, respectively, by the sets $\{a_{1},\ldots,a_{k}\}$ and $\{b_{1},\ldots,b_{k}\}$.

In turn, Duplicator has a winning strategy in this game if, and only if, there exists a homomorphism $f\colon \Ek(A)\to B$. Just observe that specifying such an $f$ is the same as specifying its coextension $f^{*}\colon \Ek(A)\to\Ek(B)$. In turn, a map $\Ek(A)\to\Ek(B)$ can be regarded as a Duplicator strategy in the existential positive game (associating with a sequence of Spoiler's moves the corresponding sequence of Duplicator's responses), and the winning condition amounts to saying that this map is a homomorphism.

\begin{remark}\label{rem:eq-elim-ep}
There is a subtlety here: suppose a homomorphism $\Ek(A)\to\Ek(B)$ sends a sequence $[a_{1},a_{2},a_{1}]$ to $[b_{1},b_{2},b_{3}]$ with $b_{1}\neq b_{3}$. This prevents the relation $\{(a_{1},b_{1}),(a_{2},b_{2}),(a_{1},b_{3})\}$ from being functional, and so it does \emph{not} define a Duplicator winning strategy. Yet, we can obtain a Duplicator winning strategy by imposing that, if Spoiler plays the same element twice (or more), Duplicator repeats their answer.

This issue can be solved by considering only the $I$-morphisms instead of all Kleisli morphisms.
In fact, there is a \emph{bijection} between $I$-morphisms $\Ek(A)\to B$ (equivalently, Kleisli morphisms $\Ek^{=}(A)\to B$) and Duplicator winning strategies in the existential positive $k$-round \EF~game.

In logical terms, the reason why Theorem~\ref{th:ep-Kleisli-morphisms} holds in the form stated, with no need to consider the comonads $\G^{=}$, is that the existential positive fragments admit equality elimination. For example, in $\EPFO_{k}$ the only atomic (sub)formulas using the equality symbol are those of the form $x = y$ with $x, y$ variables, and these can be eliminated by substituting one variable for the other in the appropriate way. 
A similar reasoning applies to the pebbling comonad. 
\end{remark}

Let us mention that the existential positive variant of pebble games was first studied in \cite{KV1995} where it is called ``existential pebble game'', while the existential positive variant of bisimulation games is known as the \emph{simulation game}.

\subsection{Fragments with counting quantifiers}

In \S\ref{s:epf}, we saw that the homomorphism preorder in the Kleisli categories for game comonads captures preservation of existential positive fragments. We shall now look at another natural relation between objects of the Kleisli categories, namely the isomorphism relation.

If $\LL$ is a class of first-order formulas, we write \[\LL(\#)\] for the extension of $\LL$ with \emph{counting quantifiers}. That is, for each $i\in\N$ we add a quantifier $\exists_{\geq i}$ such that a sentence $\exists_{\geq i} x.\phi(x)$ is satisfied in a structure $A$ if, and only if, there are at least $i$ elements of $A$ witnessing the fact that $\phi(x)$ is satisfiable in $A$. 
In the case of modal formulas, we add \emph{graded modalities} $\Diamond_{\alpha}^{i}$ such that $\Diamond_{\alpha}^{i}\phi$ holds in a pointed Kripke model $(A,a)$ if, and only if, the are at least $i$ elements of $A$ that are accessible from $a$ through $R_{\alpha}$ and satisfy $\phi$.

We will restrict ourselves to the class of finite structures. In this case, the isomorphism relation in the Kleisli category captures equivalence in the counting extensions of the logic fragments:
\begin{theorem}[\cite{abramsky2017pebbling,AS2021}]\label{t:counting-iso-kleisli}
The following statements hold for each $k,n\in\N$ and finite (pointed) structures $A,B$:
\begin{enumerate}
\item $A\equiv_{\FO_{k}(\#)} B$ if, and only if, there is an isomorphism $A\cong B$ in $\KL(\Ek^{=})$.
\item $A\equiv_{\FO^{n}(\#)} B$ if, and only if, there is an isomorphism $A\cong B$ in $\KL(\Pn^{=})$.
\item $(A,a)\equiv_{\ML_{k}(\#)} (B,b)$ if, and only if, there is an isomorphism $(A,a)\cong (B,b)$ in $\KL(\Mk)$.
\end{enumerate}
\end{theorem}

The previous result makes use of the fact that equivalence in the logics with counting quantifiers $\FO_{k}(\#)$, $\FO^{n}(\#)$ and $\ML_{k}(\#)$ are known to be captured by variants of the usual \EF, pebble and bisimulation games that consider \emph{bijections} between structures. 

For example, the \emph{bijective \EF~game} \cite{Hella1996} played between finite structures~$A$ and~$B$ proceeds as follows. If $A$ and $B$ have different cardinalities, Duplicator loses the game. Otherwise, at round $i$, Duplicator choses a bijection ${f_{i}\colon A\to B}$ and Spoiler responds by choosing an element $a_{i}\in A$. After $k$-rounds, Duplicator wins if the relation $\{(a_{i},f_{i}(a_{i}))\mid 1\leq i\leq k\}$ is a partial isomorphism. We have $A\equiv_{\FO_{k}(\#)} B$ if, and only if, Duplicator has a winning strategy in this $k$-round game. In turn, such a winning strategy induces arrows $A\to B$ and $B\to A$ in $\KL(\Ek)$ that are inverse to each other and, conversely, any isomorphism between $A$ and $B$ in $\KL(\Ek)$ induces a Duplicator winning strategy.

Similarly, the fragments $\FO^{n}(\#)$ and $\ML_{k}(\#)$ are captured, respectively, by \emph{bijective $n$-pebble games} \cite{Hella1996} and $k$-round \emph{graded bisimulation games} (cf.\ e.g.\ \cite[\S 5.3]{AS2021} or \cite{Otto2019arxiv}). Let us point out that, in the case of the pebbling comonad, we do not need to consider the infinitary extension of $\FO^{n}(\#)$ with arbitrary conjunctions and disjunctions because, as with classical pebble games, the bijective $n$-pebble game between \emph{finite} structures is determined by a $k$-round strategy for $k$ large enough.

\section{Coalgebras and combinatorial parameters}
\label{s:coalgebras}

\subsection{Coalgebras, forest covers and combinatorial parameters of structures}
\label{s:coalg-forest}

In this section, we consider the categories of Eilenberg--Moore coalgebras for game comonads and show that they encode in a structural way combinatorial parameters of structures.
We start by recalling the notion of coalgebra for a comonad; the passage from the Kleisli category of a comonad to its Eilenberg--Moore category can be understood, dually, as an abstraction of the passage from free groups to all groups. 
\begin{definition}
Let $\G$ be a comonad on a category $\C$. A \emph{coalgebra for $\G$} (also called a \emph{$\G$-coalgebra}) is a pair $(A,\alpha)$ where $A$ is an object of $\C$ and $\alpha$ is a \emph{coalgebra structure on~$A$}, i.e.\ an arrow $\alpha\colon A\to \G A$ making the following diagrams commute:
\begin{center}
\begin{tikzcd}
A \arrow{r}{\alpha} \arrow{d}[swap]{\alpha} & \G A \arrow{d}{\delta_{A}} \\
\G A \arrow{r}{\G\alpha} & \G\G A
\end{tikzcd}
\ \ \ \ \ \ \ 
\begin{tikzcd}
A \arrow{r}{\alpha} \arrow{dr}[swap]{\id_{A}} & \G A \arrow{d}{\epsilon_{A}} \\
& A
\end{tikzcd}
\end{center}

A \emph{coalgebra morphism} $(A,\alpha)\to (B,\beta)$ is an arrow $h\colon A\to B$ in $\C$ compatible with the coalgebra structures, meaning that $\beta \circ h = \G h\circ \alpha$. The \emph{Eilenberg--Moore category of~$\G$}, denoted by $\EM(\G)$, is the category consisting of $\G$-coalgebras and their morphisms. 
\end{definition}

There is an obvious forgetful functor $L\colon \EM(\G)\to\C$ that forgets the coalgebra structures. In the converse direction, there is a functor $F\colon \C\to\EM(\G)$ that sends an object~$A$ of~$\C$ to the \emph{co-free coalgebra $(\G A, \delta_{A})$ over $A$}. These functors form an adjunction
\begin{equation}\label{eq:EM-adjunction}
\begin{tikzcd}[column sep=0.8em]
\C \arrow[hbend left]{rr}{F} & \text{\footnotesize{$\top$}} & \EM(\G) \arrow[hbend left]{ll}{L}
\end{tikzcd}
\end{equation}
such that $L\circ F = \G$; this is the \emph{Eilenberg--Moore adjunction} associated with $\G$.
Moreover, the Kleisli category $\KL(\G)$ is isomorphic to the full subcategory of $\EM(\G)$ defined by the co-free coalgebras, via the functor sending a Kleisli arrow $f\colon \G A \to B$ to its coextension $f^{*}\colon \G A \to \G B$. 

\vspace{1em}
An important feature of game comonads is that they are defined uniformly with respect to the resource parameters; in fact, they are \emph{graded comonads}, with grades provided by the monoid $\N\cup\{\infty\}$ with the $\min$ operation. In particular, the resource-indexing allows us to define a notion of \emph{coalgebra number}:

\begin{definition}
Consider a sequence of comonads $(\G_{k})_{k\in\N}$ on a category $\C$. For each object $A$ of $\C$, the \emph{coalgebra number} of $A$, if it exists, is the least $\ell\in\N$ such that~$A$ admits a $\G_{\ell}$-coalgebra structure $\alpha\colon A\to \G_{\ell}A$.
\end{definition}

\subsubsection{Coalgebras for the modal comonad}\label{s:coalgebras-modal}
The modal comonads $\Mk$ are idempotent, hence the Eilenberg--Moore categories $\EM(\Mk)$ can be identified with full (coreflective) subcategories of $\CSstar$. 
Intuitively, this means that coalgebra structures for $\Mk$ are a property of Kripke models, rather than extra structure; in particular, a Kripke model admits at most one $\Mk$-coalgebra structure.
To make this precise, we recall the following notion.

\begin{definition}
A \emph{synchronization tree} is a pointed Kripke model $(A,a)\in\CSstar$ such that, for each $a'\in A$ distinct from $a$, there is a unique finite path from $a$ to $a'$:
\[
a \xrightarrow{R_{1}} a_{1} \xrightarrow{R_{2}} a_{2} \xrightarrow{\phantom{R_{3}}} \cdots \xrightarrow{R_{\ell}} a'.
\]
If each such path has length at most $k$, we say that the synchronization tree has \emph{height at most $k$}.
\end{definition}

If $(A,a)$ is a synchronization tree of height at most $k$, we can define a coalgebra structure $\alpha\colon (A,a)\to \Mk(A,a)$ as follows: 
\begin{itemize}
\item $\alpha(a)$ is the one-element sequence $[a]$, and
\item for each $a'\in A$ distinct from $a$, $\alpha(a')$ is the unique path from $a$ to $a'$.
\end{itemize}  
In fact, it turns out that a Kripke model admits an $\Mk$-coalgebra structure if, and only if, it a synchronization tree of height at most $k$. This leads to the following characterisation of the Eilenberg--Moore categories $\EM(\Ek)$:

\begin{theorem}\label{t:coalgebras-Mk}
$\EM(\Mk)$ is isomorphic to the full subcategory of $\CSstar$ consisting of synchronization trees of height at most $k$. 
\end{theorem}

The Eilenberg--Moore adjunction in eq.~\eqref{eq:EM-adjunction} specialises to an adjunction
\begin{equation*}
\begin{tikzcd}[column sep=0.8em]
\CSstar \arrow[hbend left]{rr}{F} & \text{\footnotesize{$\top$}} & \EM(\Mk). \arrow[hbend left]{ll}{L}
\end{tikzcd}
\end{equation*}
By Theorem~\ref{t:coalgebras-Mk}, we can identify $L$ with the inclusion of the subcategory of synchronization trees of height $\leq k$, and $F$ sends a pointed Kripke model to its $k$-unravelling.

Another way of looking at this is to say that the coalgebra structures are ``definable'' and thus preserved by homomorphisms of Kripke models. To make these definable structures explicit, we introduce the following terminology.

\begin{definition}\label{d:forest}
Let $(X,\leq)$ be a poset.
\begin{itemize}
\item We say that $(X,\leq)$ is a \emph{forest} if, for each $x\in X$, the down-set
\[
\down x\coloneqq \{x'\in X\mid x'\leq x\}
\]
is finite and linearly ordered by $\leq$. A minimal element of a forest is called a \emph{root}. A forest with exactly one root is a \emph{tree}.
\item If $(X,\leq)$ is a forest, the \emph{height} of an element $x\in X$ is $\htf(x)\coloneqq |\down x|-1$, and the height of $(X,\leq)$ is $\sup{\{\htf(x)\mid x\in X\}}$.
\item A \emph{forest morphism} is a monotone map between forests that preserves the height of elements.
\end{itemize}
\end{definition}

\begin{remark}\label{rem:equiv-forest-tree}
The category of forests and forest morphisms between them is equivalent to its full subcategory defined by the \emph{non-empty} trees. In one direction, a forest is sent to the tree obtained by adding a new root, and in the other direction we remove the unique root of a non-empty tree.
\end{remark}

Given a synchronization tree $(A,a)$, consider the binary relation $\cvr$ on $A$ defined, for each $x,y\in A$, by
\begin{equation}\label{eq:M}\tag{\textnormal{M}}
x\cvr y \ \Longleftrightarrow \ R_{\alpha}(x,y) \ \text{ for a unique $R_{\alpha}\in\sigma$}.
\end{equation}
The reflexive transitive closure $\leq$ of $\cvr$ is a tree order on $A$ whose root is $a$; in fact,~$\cvr$ is the \emph{covering relation} associated with $\leq$.\footnote{\label{footn:covering-rel}Given a partial order $\leq$, the associated covering relation $\cvr$ is defined by $x\cvr y$ if, and only if, $x<y$ and there is no $z$ such that $x<z<y$.} From this standpoint, an $\Mk$-coalgebra structure on a pointed Kripke model $(A,a)$ is the same thing as a tree order $\leq$ of height at most $k$ on $A$ such that $a$ is the root of the tree and the associated covering relation~$\cvr$ satisfies condition~\eqref{eq:M} above. Clearly, there is at most one such tree order on $A$. Moreover, homomorphisms of Kripke models preserve these tree orders; i.e., a homomorphism of Kripke models between synchronization trees is a forest morphism with respect to the corresponding tree orders.

In the case of the modal comonads, the coalgebra number is simply the height of a synchronization tree.

\subsubsection{Coalgebras for the \EF~comonad}
The \EF~comonads are not idempotent, so coalgebras for $\Ek$ carry extra structure that need not be preserved by homomorphisms of relational structures. As suggested by the case of coalgebras for the modal comonads, this extra structure consists of a ``compatible'' forest order. We thus define the category of \emph{forest-ordered structures}, denoted by \[\R(\sigma).\] Its objects are pairs $(A,\leq)$ where $A\in\CS$ and $\leq$ is a forest order on $A$, and morphisms are homomorphisms of $\sigma$-structures that are also forest morphisms.

In order to specify the relevant notion of compatibility between the forest order and the relations on a structure, it is convenient to use the concept of Gaifman graph:
\begin{definition}
The \emph{Gaifman graph} associated with a structure $A\in \CS$ is the undirected graph $\mathfrak{G}(A)$ defined as follows:
\begin{itemize}
\item the vertices of $\mathfrak{G}(A)$ are the elements of $A$;
\item vertices $a,b$ are adjacent in $\mathfrak{G}(A)$ if they are distinct and appear in a tuple of related elements in $A$.
\end{itemize}
\end{definition}

Given a forest-ordered structure $(A,\leq)$, the compatibility condition for $\Ek$-coalgebras is the following:
\begin{equation}\label{eq:E}\tag{\textnormal{E}}
\forall a,b\in A, \text{ if $a,b$ are adjacent in $\mathfrak{G}(A)$ then they are comparable in the order $\leq$}.
\end{equation}

\begin{theorem}\label{t:coalgebras-Ek}
$\EM(\Ek)$ is isomorphic to the full subcategory of $\R$ defined by the forest-ordered structures of height at most $k$ satisfying condition~\eqref{eq:E}.
\end{theorem}

The previous result entails that, in the Eilenberg--Moore adjunction 
\begin{equation*}
\begin{tikzcd}[column sep=0.8em]
\CS \arrow[hbend left]{rr}{F} & \text{\footnotesize{$\top$}} & \EM(\Ek) \arrow[hbend left]{ll}{L}
\end{tikzcd}
\end{equation*}
obtained by specialising eq.~\eqref{eq:EM-adjunction}, we can identify $L$ with the functor forgetting the forest order, while $F$ sends a structure $A$ to the forest-ordered structure $(\Ek(A),\sqsubseteq)$, where~$\sqsubseteq$ is the prefix order. In contrast to the modal case, the forest order of an \EF~coalgebra is not determined by its underlying relational structure.

Theorem~\ref{t:coalgebras-Ek} tells us, in particular, that $\Ek$-coalgebra structures on $A\in\CS$ are in bijection with \emph{forest covers} of $A$ (i.e., forest orders on $A$ satisfying~\eqref{eq:E}) of height at most $k$. The least height of a forest cover of $A$, referred to as the \emph{tree-depth} of $A$, is an important combinatorial parameter of structures which was first introduced for graphs in~\cite{nevsetvril2006tree}. Thus, an $\Ek$-coalgebra structure on~$A$ is a witness to the fact that the tree-depth of $A$ is at most~$k$. This yields:

\begin{proposition}\label{p:coalg-numb-Ek}
For the \EF~comonads $\Ek$, the coalgebra number of a structure $A\in\CS$ coincides with the tree-depth of $A$.
\end{proposition}

\subsubsection{Coalgebras for the pebbling comonad}
To characterise the categories of Eilenberg--Moore coalgebras for the pebbling comonads $\Pn$, we consider a slight variant of the category $\R$ of forest-ordered structures that takes into account the presence of pebbles. For each positive integer $n$, we define the category \[\R^{(n)}\] of \emph{$n$-pebble forest-ordered structures}: the objects are tuples $(A,\leq,p)$ such that $(A,\leq)$ is a forest-ordered structure and $p\colon A\to \n$ is a map, called \emph{pebbling function}, assigning to each element of $A$ an integer between $1$ and $n$; the morphisms in $\R^{(n)}$ are the morphisms of forest-ordered structures that also preserve the pebbling functions.

For an $n$-pebble forest-ordered structure $(A,\leq,p)$, the compatibility between the order and the pebbling function is captured by the following condition:
\begin{equation}\label{eq:P}\tag{\textnormal{P}}
\begin{split}
\forall a,b\in A, \ \ &\text{if $a,b$ are adjacent in $\mathfrak{G}(A)$ and $a<b$ in the forest order}, \\ &\text{then $p(a)\neq p(x)$ for all $x\in A$ such that $a<x\leq b$}.
\end{split}
\end{equation}
\begin{theorem}\label{t:coalgebras-Pn}
$\EM(\Pn)$ is isomorphic to the full subcategory of $\R^{(n)}$ defined by the $n$-pebble forest-ordered structures satisfying conditions~\eqref{eq:E} and~\eqref{eq:P}.
\end{theorem}

The previous result allows us to give a concrete description of the Eilenberg--Moore adjunction between $\CS$ and $\EM(\Pn)$: the left adjoint $L\colon \EM(\Pn)\to \CS$ forgets both the forest orders and the pebbling functions, whereas the right adjoint $F\colon \CS\to \EM(\Pn)$ sends a structure $A$ to $(\Pn(A),\sqsubseteq, p)$ where the pebbling function $p\colon \Pn(A)\to \n$ sends a sequence $[(p_{1},a_{1}), \ldots, (p_{\ell},a_{\ell})]$ to $p_{\ell}$.

It follows from Theorem~\ref{t:coalgebras-Pn} that $\Pn$-coalgebra structures on $A\in\CS$ are in bijection with \emph{$n$-pebble forest covers} of $A$, i.e.\ pairs of forest orders and pebbling functions on $A$ satisfying~\eqref{eq:E} and~\eqref{eq:P}. Furthermore, it turns out that $A$ admits an $n$-pebble forest cover if, and only if, it admits a \emph{tree decomposition of width $<n$} in the sense of \cite{robertson1986graph,FV1998}; cf.~\cite{abramsky2017pebbling} or \cite[Theorem~6.4]{AS2021}. The least width of a tree decomposition of $A$ is called the \emph{tree-width} of $A$, a combinatorial parameter that plays an important role in finite model theory. Thus, $\Pn$-coalgebra structures on $A$ witness the fact that the tree-width of $A$ is $< n$:

\begin{proposition}\label{p:coalg-numb-Pn}
For the pebbling comonads $\Pn$, the coalgebra number of a structure $A\in\CS$ coincides with the tree-depth of $A$ increased by $1$.\footnote{The definition of width of a tree decomposition customarily involves subtracting $1$, so that trees have tree-width $1$; this is why there is a difference of $1$ between coalgebra numbers and tree-width.}
\end{proposition}

\subsection{Homomorphism counting results \emph{\`a la} Lov\'asz}
A well-known consequence of the \emph{Yoneda Lemma} states that any two objects $X,Y$ of an (essentially small) category $\C$ are isomorphic if, and only if, the corresponding hom-functors are \emph{naturally isomorphic}; in symbols,
\[
X\cong Y \ \Longleftrightarrow \ \hom(-,X)\cong \hom(-,Y).
\]
In 1967, Lov\'asz showed that the naturality condition can be omitted when $\C$ is the category of finite $\sigma$-structures, for any finite relational signature $\sigma$ \cite{Lovasz1967}. That is, finite $\sigma$-structures $A,B$ are isomorphic precisely when they admit the same (finite) number of homomorphisms from any other finite $\sigma$-structure:
\[
A\cong B \ \Longleftrightarrow \ \text{for all finite $\sigma$-structures $C$, } \hom(C,A)\cong \hom(C,B).
\]

\begin{remark}
Two pointwise isomorphic functors (as in Lov\'asz' theorem) need not be naturally isomorphic, as the next example from \cite{Joyal1981} shows. For any finite set~$X$, let $K_{1}(X)$ and $K_{2}(X)$ be, respectively, the set of permutations of~$X$ and the set of linear orders on~$X$. The assignments $X\mapsto K_{1}(X)$ and $X\mapsto K_{2}(X)$ can be extended to endofunctors~$K_{1}$ and~$K_{2}$, respectively, on the category of finite sets and bijections between them. The functors~$K_{1}$ and~$K_{2}$ are pointwise isomorphic because the number of permutations of a finite set ~$X$ is~$|X|!$, which is also the number of linear orders on~$X$. However, they are not naturally isomorphic; just note that there is no natural way of assigning a linear order to the identity permutation.
\end{remark}

Homomorphism counting results, akin to Lov\'asz' seminal result, have been investigated also in finite model theory. There, one is interested in equivalence relations between structures that can be captured in terms of homomorphism counts from a restricted class of structures. Prime examples are Dvo\v{r}\'ak's and Grohe's homomorphism counting results:

\begin{theorem}[\cite{dvovrak2010recognizing}]
Let $\C$ be the category of finite undirected graphs without loops and graph homomorphisms, and let $\mathcal{K}\subseteq \C$ consist of those graphs having tree-width $<n$. For any $A,B\in \C$ we have
\[
A\equiv_{\FO^{n}(\#)} B \ \Longleftrightarrow \ \text{for all $C\in\mathcal{K}$, } \hom(C,A)\cong \hom(C,B).
\]
\end{theorem}

\begin{theorem}[\cite{grohe2020counting}]
Let $\C$ be the category of finite undirected vertex-coloured graphs and graph homomorphisms preserving the colours, and let $\mathcal{K}\subseteq \C$ consist of those graphs having tree-depth $\leq k$. For any $A,B\in \C$ we have
\[
A\equiv_{\FO_{k}(\#)} B \ \Longleftrightarrow \ \text{for all $C\in\mathcal{K}$, } \hom(C,A)\cong \hom(C,B).
\]
\end{theorem}

The three homomorphism counting results above find a natural common generalisation in the setting of game comonads. Recall from Theorem~\ref{t:counting-iso-kleisli} that equivalence in the counting logics is captured by the isomorphism relation in the Kleisli categories, or equivalently in the Eilenberg--Moore categories (since the former fully embed into the latter). In turn, by Propositions~\ref{p:coalg-numb-Ek} and~\ref{p:coalg-numb-Pn}, the structures having tree-width $< n$ or tree-depth $\leq k$ are precisely those admitting, respectively, a coalgebra structure for $\Pn$ or $\Ek$. This suggests the following notion:

\begin{definition}
Let $\C$ be a locally finite category (that is, there are finitely many arrows between any two objects of $\C$). We say that $\C$ is \emph{combinatorial} if, for all $A,B\in \C$,
\[
A\cong B \ \Longleftrightarrow \ \text{for all $C\in \C$, } \hom(C,A)\cong \hom(C,B).
\]
\end{definition}

For example, Lov\'asz' theorem states that the category of finite $\sigma$-structures is combinatorial. In fact, any locally finite category with pushouts and a proper factorisation system is combinatorial; see~\cite{DJR2021} and also \cite{pultr1973isomorphism,isbell1991some}. (The assumption that all pushouts exist is not necessary and can be weakened, see \cite[Remark~13]{DJR2021} and \cite[Theorem~4.3]{Reggio2022}.) This leads to the following result:

\begin{theorem}[\cite{DJR2021}]\label{t:Lovasz-abstract-coalgebras}
Let $\G$ be a comonad on $\CS$. The full subcategory of $\EM(\G)$ consisting of the finite coalgebras is combinatorial.\footnote{A coalgebra for $\G$ is \emph{finite} if its underlying $\sigma$-structure is finite.}
\end{theorem}

If we let $\G$ in Theorem~\ref{t:Lovasz-abstract-coalgebras} be the identity comonad, we obtain exactly Lov\'asz' theorem. When $\G=\Ek^{I}$, in view of Remark~\ref{rem:comparison-functor-Kleisli} and Theorem~\ref{t:counting-iso-kleisli} we get that any two finite $\sigma$-structures $A$ and~$B$ are $\FO_{k}(\#)$-equivalent if, and only if, $J(A)$ and $J(B)$ cannot be distinguished by the number of homomorphisms from any finite $\sigma^{I}$-structure of tree-depth $\leq k$. One then shows that the functor $H\colon \CSplus\to\CS$, left adjoint to $J$, does not increase the tree-depth of structures, and so $A$ and $B$ are $\FO_{k}(\#)$-equivalent if, and only if, they cannot be distinguished by the number of homomorphisms from any finite $\sigma$-structure of tree-depth $\leq k$. 

A similar analysis can be carried out for the pebbling comonad~$\Pn^{I}$, and thus we obtain variants of Dvo\v{r}\'ak's and Grohe's theorems for arbitrary structures.\footnote{The case of the pebbling comonad is less straightforward because, unlike for the \EF~comonad, co-free coalgebras for the pebbling comonad are never finite. To overcome this hurdle, one can either ``stratify'' the pebbling comonad into finitary levels or apply a more direct topological argument; see \cite[{\S}IV.B]{DJR2021} and \cite[\S6.3]{Reggio2022}, respectively.} The aforementioned results, in their original form, can be recovered simply by observing that the comonads restrict to the appropriate categories of graphs; cf.\ \cite{DJR2021}. 

The analysis outlined above, and in particular the observation that the functor $H$ preserves the relevant combinatorial parameters of structures, has the following logical consequences concerning equality elimination:

\begin{corollary}[\cite{DJR2021}]
Let $\LL$ be either $\FO_{k}(\#)$ or $\FO^{n}(\#)$. The following statements hold:
\begin{enumerate}
\item If two finite $\sigma$-structures are not distinguished by any sentence of $\LL$ without equality, then they are not distinguished by any sentence of $\LL$.
\item Let $\LL_{\infty}$ be the closure of $\LL$ under infinitary conjunctions and disjunctions. Each sentence of $\LL_{\infty}$ is equivalent, over finite $\sigma$-structures, to one without equality.
\end{enumerate}
\end{corollary}

Finally, let us mention that applying (the obvious variant for pointed structures of) Theorem~\ref{t:Lovasz-abstract-coalgebras} to the modal comonad, one obtains at once a homomorphism counting result for graded modal logic that characterises $\ML_{k}(\#)$-equivalence of finite pointed Kripke models in terms of homomorphism counts from finite synchronization trees of height $\leq k$ \cite[Theorem~31]{DJR2021}.

\section{Paths, open maps, and more logical equivalences}
\label{s:open}

In \S\ref{s:logical-equiv} we saw that the homomorphism preorder and the isomorphism relation in the Kleisli categories for game comonads capture, respectively, preservation of existential positive fragments $\exists^{+}\LL$ and equivalence (of finite structures) in fragments with counting quantifiers $\LL(\#)$. 
Note that, since the Kleisli category $\KL(\G)$ of a comonad~$\G$ is a full subcategory of the Eilenberg--Moore category $\EM(\G)$, these results could equivalently be phrased in the latter category. 

In this section, we show how other relations between coalgebras for game comonads, which are \emph{not} definable in the Kleisli category, allow us to encode equivalence in the \emph{full fragments} $\LL$ as well as preservation of the \emph{existential fragments} $\exists\LL$ and \emph{positive fragments}~${}^{+}\LL$. 
These results crucially rely on the notion of \emph{path}. We give a concrete definition of path using the characterisation of Eilenberg--Moore categories of game comonads in \S\ref{s:coalgebras}; an axiomatic definition of path will be given in \S\ref{s:arboreal}.

\begin{definition}\label{d:concrete-path}
A coalgebra $(A,\alpha)$ for any of the comonads $\Ek$, $\Pn$ or $\Mk$ is a \emph{path} if, when regarded as a forest-ordered structure $(A,\leq)$, it is a finite linear order. Paths will be denoted by $P, Q$ and variations thereof.
\end{definition}

Before we continue, let us make the following point about the handling of the equality symbol. Recall from \S\ref{s:equality} the comonads $\G^{I}$ and $\G^{=}$, for a generic game comonad $\G$ on $\CS$. We have seen that there is a full embedding $\KL(\G^{=})\to \KL(\G^{I})$ sending $A$ to $J(A)$ (see Remark~\ref{rem:comparison-functor-Kleisli}), and therefore the existence of an (iso)morphism $A\to B$ in $\KL(\G^{=})$ is equivalent to the existence of an (iso)morphism $J(A)\to J(B)$ in $\KL(\G^{I})$.

When considering the Eilenberg--Moore categories for the comonads $\G^{I}$ and $\G^{=}$, the situation is quite different. There is a canonical comparison functor 
\[
\EM(\G^{I})\to \EM(\G^{=}),
\]
but it is \emph{not} fully faithful (this follows e.g.\ from the criterion dual to the one in \cite[Proposition~2.2]{Mesablishvili2006}). Since the category $\EM(\G^{I})$ typically admits a concrete and simple description as outlined in \S\ref{s:coalg-forest}, instead of working with (the Eilenberg--Moore adjunction associated with) the comonad $\G^{=}$, we work with its \emph{resolution}
\begin{equation}\label{eq:I-adjunction}
\begin{tikzcd}[column sep=0.8em]
\CS \arrow[hbend left]{rr}{J} & \text{\footnotesize{$\top$}} & \CSplus \arrow[hbend left]{rr}{F^{I}} \arrow[hbend left]{ll}{H} & \text{\footnotesize{$\top$}} & \EM(\G^{I}) \arrow[hbend left]{ll}{L^{I}}
\end{tikzcd}
\end{equation}
where $L^{I}\dashv F^{I}$ is the Eilenberg--Moore adjunction associated with $\G^{I}$. In order to improve readability, we will use the notation
\begin{equation}\label{eq:FI}
\FI\coloneqq F^{I}\circ J \colon \CS\to \EM(\G^{I}).
\end{equation}

\subsection{Existential fragments and pathwise embeddings}\label{s:exist-pe}

Given a class of first-order formulas $\LL$, we write 
\[
\exists\LL
\]
for the fragment of $\LL$ consisting of \emph{existential formulas}, i.e.\ those formulas that contain no universal quantifiers and in which negations are only applied to atomic subformulas. In the case of modal logic, the existential formulas are those in which the modalities $\Box_{\alpha}$ do not appear and the negation symbol is only applied to propositional variables. Since preserving the validity of negated atomic formulas is equivalent to \emph{reflecting} the validity of atomic formulas, we should look at coalgebra morphisms that not only preserve relations but also reflect them (at least in appropriate cases). This leads us to the notion of pathwise embedding.

Recall that a homomorphism of $\sigma$-structures $h\colon A\to B$ is an \emph{embedding} if it is injective and, for each relation symbol $R\in \sigma$ of arity $n$ and each $(a_{1},\ldots,a_{n})\in A^{n}$, 
\[
(h(a_{1}),\ldots,h(a_{n}))\in R^{B} \ \Longrightarrow \ (a_{1},\ldots,a_{n})\in R^{A}.
\]
Now, let $\A$ be any of the categories $\EM(\Ek)$, $\EM(\Pn)$ or $\EM(\Mk)$. 
\begin{definition}\label{d:concrete-pathwise-emb}
Let $f\colon X\to Y$ be an arrow in $\A$.
\begin{itemize}
\item $f$ is an \emph{embedding}, written $\emb$, if its underlying homomorphism of $\sigma$-structures is.
\item If $X$ is a path and $f$ is an embedding, then $f$ is called a \emph{path embedding}.
\item $f$ is a \emph{pathwise embedding} if, for all path embeddings $m\colon P\emb X$, the composite $f\circ m\colon P\to Y$ is a path embedding. That is, the restriction of $f$ to each branch of the forest is an embedding.
\end{itemize}
\end{definition}

\begin{remark}
Every embedding in $\A$ is a pathwise embedding; the converse need not hold because pathwise embeddings may fail to be injective.
\end{remark}

Akin to the homomorphism preorder, we can consider the preorder between objects determined by the existence of a pathwise embedding between them. The next result states that this preorder captures exactly preservation of the existential fragments. Recall the functor $\FI$ from eq.~\eqref{eq:FI}; with this notation, we have
\begin{theorem}\label{th:e-pathwise-emb}
The following statements hold for each $k,n\in\N$ and (pointed) structures $A,B$: 
\begin{enumerate}
\item $A\IMP_{\EFO_{k}} B$ if and only if there is a pathwise embedding $\FI (A)\to \FI(B)$ in $\EM(\Ek^{I})$.
\item $A\IMP_{\exists L^{n}_{\infty,\omega}} B$ if and only if there is a pathwise embedding ${\FI (A)\to \FI(B)}$ in $\EM(\Pn^{I})$.
\item\label{i:exist-pe-modal} $(A,a)\IMP_{\EML_{k}} (B,b)$ if and only if there is a pathwise embedding $F(A,a)\to F(B,b)$ in $\EM(\Mk)$.
\end{enumerate}
\end{theorem}

While Theorem~\ref{th:ep-Kleisli-morphisms} relied on the correspondence between Kleisli arrows and Duplicator winning strategies in the existential positive variant of the games, Theorem~\ref{th:e-pathwise-emb} relies on a correspondence between pathwise embeddings and Duplicator winning strategies in the \emph{existential} variant of the games. In these games, Spoiler is forced to play in $A$ and Duplicator responds in $B$, and the winning condition is the same as in the back-and-forth game.

\subsection{Full fragments and open maps}

In order to capture equivalence of structures in the full fragments $\LL$, corresponding to equivalence in the back-and-forth games, we consider those pathwise embeddings that satisfy the following path lifting property (as before, $\A$ denotes any of the categories $\EM(\Ek)$, $\EM(\Pn)$ or $\EM(\Mk)$):

\begin{definition}\label{d:open-concrete}
A morphism $f\colon X\to Y$ in $\A$ is \emph{open} if each commutative square
\[\begin{tikzcd}
P \arrow[rightarrowtail]{r} \arrow[rightarrowtail]{d} & X \arrow{d}{f} \\
Q \arrow[rightarrowtail]{r} \arrow[rightarrowtail,dashed]{ur} & Y
\end{tikzcd}\]
where the horizontal morphisms and the left vertical one are path embeddings, admits a \emph{diagonal filler}, i.e.\ an arrow $Q\to X$ making the two triangles commute (such an arrow is necessarily an embedding).
\end{definition}

We use the notation $\open$ to indicate that a morphism is an open pathwise embedding.
Given objects $X,Y$ of $\A$, a \emph{bisimulation} between $X$ and $Y$ is a span of open pathwise embeddings connecting the two objects:
\[
X \openleft W \open Y
\]
If such a bisimulation exists, we say that $X$ and $Y$ are \emph{bisimilar} and write $X\lr Y$.

\begin{remark}
An arrow $f\colon X\to Y$ in $\A$ is an open pathwise embedding if, and only if, for all paths $P,Q$, each commutative square
\[\begin{tikzcd}
P \arrow{r} \arrow{d} & X \arrow{d}{f} \\
Q \arrow{r} \arrow[dashed]{ur} & Y
\end{tikzcd}\]
admits a diagonal filler \cite[Lemma~4.1]{AR2022}. Thus, the open pathwise embeddings are precisely the arrows in $\A$ that satisfy the \emph{right lifting property} with respect to all morphisms between paths (cf.\ \S\ref{ss:arboreal}).
\end{remark}

To see how bisimulations yield winning strategies in back-and-forth games, suppose there exists a bisimulation 
\[
X \mathrel{\mathop{\openleft}^{f}} W \mathrel{\mathop{\open}^{g}} Y.
\] 
We consider a back-and-forth game between $X$ and $Y$ whose positions are pairs of path embeddings
\[
(P\emb X, P\emb Y)
\]
with the same domain. Assume that the current position is given by
\[
(f\circ m\colon P\emb X, g\circ m\colon P\emb Y)
\]
for some path embedding $m\colon P\emb W$.
In the next round, Spoiler plays a path embedding ${j\colon Q\emb X}$ extending $f\circ m$, and Duplicator must respond with a path embedding $k\colon Q\emb Y$ extending $g\circ m$. Duplicator can use the fact that $f$ is an open pathwise embedding to obtain a diagonal filler $n$ as displayed below.
\[\begin{tikzcd}
P \arrow[rightarrowtail]{r}{m} \arrow[rightarrowtail]{d} & W \arrow{d}{f} \\
Q \arrow[rightarrowtail]{r}{j} \arrow[rightarrowtail,dashed]{ur}[description]{n} & X
\end{tikzcd}\]
Duplicator's response will then be $g\circ n$, which is a path embedding since $g$ is a pathwise embedding. Symmetrically, if Spoiler plays a path embedding $j'$ into $Y$, Duplicator can use the fact that $g$ is an open pathwise embedding to ``lift $j'$ along $g$'' and then compose this lifting with $f$ to obtain a path embedding into $X$.

In fact, this back-and-forth game between objects of $\A$ can be defined axiomatically, and the existence of a Duplicator winning strategy turns out to be equivalent to the existence of a bisimulation between the objects, cf.\ \S\ref{s:arboreal}. For each of the game comonads $\Ek$, $\Pn$ and $\Mk$, the axiomatic game played between co-free coalgebras coincides with the corresponding concrete notion of model-comparison game. We thus obtain that equivalence in the full fragments is captured by the bisimilarity relation:
\begin{theorem}\label{th:full-bisim}
The following statements hold for each $k,n\in\N$ and (pointed) structures $A,B$: 
\begin{enumerate}
\item $A\equiv_{\FO_{k}} B$ if, and only if, $\FI(A)\lr \FI(B)$ in $\EM(\Ek^{I})$.
\item\label{i:ep-n-var} $A\equiv_{L^{n}_{\infty,\omega}} B$ if, and only if, $\FI(A)\lr \FI(B)$ in $\EM(\Pn^{I})$.
\item\label{i:modal-b-and-f} $(A,a)\equiv_{\ML_{k}} (B,b)$ if, and only if, $F(A,a)\lr F(B,b)$ in $\EM(\Mk)$.
\end{enumerate}
\end{theorem}

For example, an open pathwise embedding in the category $\EM(\Mk)$ is the same thing as a \emph{p-morphism}, also called \emph{bounded morphism}, a basic notion of modal logic (see e.g.\ \cite[Definition~2.10]{blackburn2002modal}). Thus, two synchronization trees are bisimilar as objects of $\EM(\Mk)$, in the sense defined above, precisely when they are bisimilar in the usual sense of modal logic (see e.g.\ \cite[Definition~2.16]{blackburn2002modal}). Therefore, item~\ref{i:modal-b-and-f} in Theorem~\ref{th:full-bisim} amounts to the well-known fact that two pointed Kripke models satisfy the same modal formulas of modal depth at most $k$ if, and only if, their $k$-unravellings are bisimilar.

Theorem~\ref{th:full-bisim} can be proved by first showing that the existence of a bisimulation is equivalent to the existence of a Duplicator winning strategy in the corresponding back-and-forth game played between $A$ and $B$, and then applying the corresponding translation result between games and logic (see Theorems~\ref{t:game-logic-modal},~\ref{th:EF} and~\ref{th:pebble}).

\subsection{Positive fragments and positive bisimulations}
If $\LL$ is a class of first-order formulas, we denote by 
\[
{}^{+}\LL
\]
the fragment of $\LL$ consisting of \emph{positive formulas}, i.e.\ formulas that do not contain the negation symbol. Akin to the way in which equivalence in the full fragments is captured by bisimulations, equivalence in the positive fragments is captured by a positive variant of bisimulations:

\begin{definition}\label{d:positive-bisim}
Let $X,Y$ be objects of $\A$. A \emph{positive bisimulation} from $X$ to $Y$ is a triple $(f,\iota,g)$
\[\begin{tikzcd}[]
W_{1} \arrow[r, "\iota"] \arrow[d,"f"', circlearrow] & W_{2} \arrow[d,"g", circlearrow] \\
X & Y
\end{tikzcd}\]
where $f$ and $g$ are open pathwise embeddings and $\iota$ is a bijection. We write $X\pb Y$ to indicate the existence of a positive bisimulation from $X$ to $Y$.
\end{definition}

The fact that $\iota$ is a bijection means that we can think of $W_{1}$ and $W_{2}$ as having the same underlying forest order, with $W_{2}$ obtained from $W_{1}$ by adding relations (i.e., by enlarging the interpretations of the relation symbols in $\sigma$). This reflects the directed nature of positive bisimulations. 

\begin{theorem}[\cite{ALR23}]\label{th:positive-pos-bisim}
The following statements hold for each $k,n\in\N$ and (pointed) structures $A,B$: 
\begin{enumerate}
\item $A\IMP_{{}^{+}\FO_{k}} B$ if, and only if, $\FI(A)\pb \FI(B)$ in $\EM(\Ek^{I})$.
\item\label{i:p-n-var} $A\IMP_{{}^{+}L^{n}_{\infty,\omega}} B$ if, and only if, $\FI(A)\pb \FI(B)$ in $\EM(\Pn^{I})$.
\item\label{i:modal-p} $(A,a)\IMP_{{}^{+}\ML_{k}} (B,b)$ if, and only if, $F(A,a)\pb F(B,b)$ in $\EM(\Mk)$.
\end{enumerate}
\end{theorem}

In the case of modal logic, our notion of positive bisimulation coincides with that of ``directed simulation'' from \cite{KdR1997}. In \emph{op.\ cit.}, the authors show that, for any two \emph{image-finite} pointed Kripke models $(A,a)$ and $(B,b)$,
\[
(A,a)\IMP_{{}^{+}\ML} (B,b) \text{ if, and only if, } (A,a)\pb (B,b).
\] 
Item~\ref{i:modal-p} in Theorem~\ref{th:positive-pos-bisim} can be regarded as a graded refinement of this result which holds for all pointed Kripke models.

Akin to Theorem~\ref{th:full-bisim}, Theorem~\ref{th:positive-pos-bisim} relies on a correspondence between positive bisimulations and Duplicator winning strategies in the \emph{positive} variant of the corresponding back-and-forth game. In this game, Spoiler can play in either $A$ or $B$, and the winning condition is the same as in the existential positive variant of the game (cf.\ \S\ref{s:epf}).

\section{An axiomatic perspective: Arboreal categories}
\label{s:arboreal}

In the previous sections we have seen that the comonads $\Ek$, $\Pn$ and $\Mk$ exhibit a similar behaviour. In each case, their coalgebras can be represented concretely as forest-ordered structures, and preservation of corresponding logic fragments can be captured in terms of natural relations between co-free coalgebras, such as the homomorphism preorder, the isomorphism and bisimilarity relations, and variations thereof. 

It is natural to ask if we can reason about a \emph{generic} game comonad. (Note, however, that there is as yet  no mathematical definition of a game comonad---you know one when you see one!) To this end, we introduce the framework of arboreal categories, which capture in an axiomatic way the main features of the categories of coalgebras for game comonads. In \S\ref{s:HPT}, we will show how this axiomatic approach can be used to give a uniform treatment of equi-resource homomorphism preservation theorems in logic.

\subsection{Arboreal categories}\label{ss:arboreal}

Within the categories of coalgebras $\EM(\Ek)$, $\EM(\Pn)$ and $\EM(\Mk)$, a class of objects that plays a key role is that of paths (see Definition~\ref{d:concrete-path}): they allow us to define pathwise embeddings, open maps, bisimulations, and so forth. It turns out that to axiomatise the concrete notion of path, as well as a considerable part of the structure of the aforementioned categories of coalgebras, all we need is a ``good'' factorisation system.

\subsubsection{Proper factorisation systems and paths}
Given arrows~$e$ and~$m$ in a category $\C$, we say that $e$ has the \emph{left lifting property} with respect to~$m$, or that~$m$ has the \emph{right lifting property} with respect to~$e$, if each commutative square
\[\begin{tikzcd}
{\cdot} \arrow{d}[swap]{e} \arrow{r} & {\cdot} \arrow{d}{m} \arrow[leftarrow, dashed]{dl} \\
{\cdot} \arrow{r} & {\cdot}
\end{tikzcd}\]
admits a diagonal filler (cf.\ Definition~\ref{d:open-concrete}). For any class $\mathscr{H}$ of morphisms in $\C$, let~$\llp{\mathscr{H}}$ (respectively, $\rlp{\mathscr{H}}$) be the class of morphisms having the left (respectively, right) lifting property with respect to all morphisms in $\mathscr{H}$.

\begin{definition}\label{def:f-s}
A \emph{weak factorisation system} in a category $\C$ is a pair of classes of morphisms $(\Q,\M)$ satisfying the following conditions:
\begin{enumerate}[label=(\roman*)]
\item Each morphism $f$ in $\C$ can be decomposed as $f = m \circ e$ with $e\in \Q$ and $m\in \M$.
\item $\Q=\llp{\M}$ and $\M=\rlp{\Q}$.
\end{enumerate}
A \emph{proper factorisation system} is a weak factorisation system satisfying the additional condition:
\begin{enumerate}[label=(\roman*), resume]
\item Every arrow in $\Q$ is an epimorphism and every arrow in $\M$ a monomorphism.
\end{enumerate}
We refer to $\M$-morphisms as \emph{embeddings} and denote them by $\emb$. $\Q$-morphisms will be referred to as \emph{quotients} and denoted by~$\epi$.
\end{definition}

Let $\C$ be a category endowed with a proper factorisation system $(\Q,\M)$.
In the same way that one usually defines the poset of subobjects of a given object $X\in\C$, we can define the poset $\Emb{X}$ of \emph{$\M$-subobjects} of $X$, whose elements are equivalence classes~$[m]$ of embeddings $m\colon S\emb X$ (whenever convenient, we denote an equivalence class $[m]$ by any of its representatives).
For any morphism $f\colon X\to Y$ and embedding $m\colon S\emb X$, we can consider the $(\Q,\M)$-factorisation  of $f\circ m$: 
\[\begin{tikzcd}
S\arrow[twoheadrightarrow]{r} & \exists_f S \arrow[rightarrowtail]{r}{\exists_{f}m} & Y.
\end{tikzcd}\] 
This yields a monotone map $\exists_f\colon \Emb{X}\to\Emb{Y}$ sending $m$ to $\exists_{f}m$, and in fact it defines a functor from $\C$ to the category of posets and monotone maps.

\begin{definition}
An object $X$ of $\C$ is called a \emph{path} provided the poset $\Emb{X}$ is a finite linear order. Paths will be denoted by $P,Q$ and variations thereof. A \emph{path embedding} is an embedding $P\emb X$ whose domain is a path. 
\end{definition}

\begin{example}\label{ex:paths-EM-categories}
In the categories $\EM(\Ek)$, $\EM(\Pn)$ and $\EM(\Mk)$, consider the proper factorisation systems $(\Q,\M)$ where~$\Q$ consists of the surjective homomorphisms, and~$\M$ consists of the embeddings in the sense of Definition~\ref{d:concrete-pathwise-emb}. In each case, the axiomatic notion of path coincides with the concrete one given in Definition~\ref{d:concrete-path}.
\end{example}

\subsubsection{Arboreal categories defined}
Now that we have an abstract notion of path, we can define path embeddings, pathwise embeddings, bisimulations, and so forth, by generalising in a straightforward manner the definitions given in the concrete setting. In particular, we can define a notion of bisimilarity in any category equipped with a proper factorisation system.

\begin{remark}
The bisimilarity relation in arboreal categories is reminiscent of (and inspired by) the notion of \emph{open-map bisimilarity} introduced in~\cite{JNW1993,JNW1996}. However, there are some differences: while the latter takes as parameter a full subcategory whose objects are called ``paths'', in the arboreal setting the notion of path is completely determined by the factorisation system (which is thus the only ``parameter''). Moreover, the arboreal notion of open morphism is a refinement of the one used in~\cite{JNW1993,JNW1996}, which in turn is a special case of the axiomatic concept of open map in toposes~\cite{JM1994}. However, the two notions of open morphisms coincide for pathwise embeddings \cite[Lemma~4.1]{AR2022}.
\end{remark}

The next step is to identify sufficient conditions on the category so that this notion of bisimilarity is well-behaved. In this regard, the sub-poset 
\[
\Path{X}
\] 
of $\Emb{X}$ consisting of the path embeddings plays a central role. In order for the maps $\exists_f\colon \Emb{X}\to\Emb{Y}$ to descend to maps $\Path{X}\to\Path{Y}$, we need that the quotient of a path is again a path. This in turn requires that the factorisation system $(\Q,\M)$ is \emph{stable}, i.e.\ for every ${e\in \Q}$ and ${m\in\M}$ with common codomain, the pullback of~$e$ along~$m$ exists and belongs to $\Q$. Together with three additional axioms ensuring that the collection of paths behaves as expected, this leads to the notion of an arboreal category:

\begin{definition}\label{def:arboreal-cat}
An \emph{arboreal category} is a category $\A$, equipped with a stable proper factorisation system, that satisfies the following conditions:
\begin{description}\itemsep4pt
\item[Paths are connected]\label{ax:connected} Coproducts of paths exist and each path $P$ is \emph{connected}, i.e.\ every arrow $P\to \coprod_{i\in I}{Q_i}$ into a coproduct of paths (indexed by a non-empty set $I$) factors through some coproduct injection. 

\item[2-out-of-3 property]\label{ax:2-out-of-3} Given arrows $f\colon P\to Q$ and $g\colon Q\to Q'$ between paths, if any two of $f$, $g$ and $g\circ f$ are quotients, then so is the third. 

\item[Path-generation]\label{i:path-gen} Each object is \emph{path-generated}, i.e.\ it is the colimit of the paths that embed into it.\footnote{\label{footn:dense-paths}More precisely, for each object~$X$, the cocone consisting of all path embeddings into~$X$, with morphisms between paths the embeddings that make the obvious triangles commute, is a colimit cocone. Equivalently, the full subcategory of $\A$ defined by the paths is \emph{dense} \cite[Theorem~5.1]{AR2022}.}
\end{description}
\end{definition}

\begin{example}\label{ex:arboreal-cats}
We give some examples of arboreal categories:
\begin{enumerate}
\item\label{i:forests-trees} The category of forests and forest morphisms is arboreal, when equipped with the factorisation system (surjective morphisms, injective morphisms). The paths are the finite chains, i.e.\ the forests consisting of a single branch. Coproducts are given by disjoint union; in particular, the initial object is the empty forest. Finite chains are easily seen to be connected, and the 2-out-of-3 property holds because forest morphisms preserve height. Finally, each forest is path-generated because it is the colimit of the diagram given by its branches and the embeddings between them.

The category $\T$ of non-empty trees and forest morphisms is equivalent to the category of forests (see Remark~\ref{rem:equiv-forest-tree}) and so it also arboreal. Paths in $\T$ are the non-empty finite chains. Non-empty coproducts in $\T$ are given by smash sum, with the roots identified, and the initial object of $\T$ is the one-element tree. 

\item The categories of coalgebras $\EM(\Ek)$, $\EM(\Pn)$ and $\EM(\Mk)$, equipped with the factorisation systems defined in Example~\ref{ex:paths-EM-categories}, are arboreal. Coproducts in $\EM(\Ek)$ and $\EM(\Pn)$ are computed in the category of forests, while coproducts in $\EM(\Mk)$ are computed in $\T$. The validity of the axioms defining arboreal categories then follows from the corresponding fact for forests and non-empty trees.

\item\label{i:presh-forest} If $X$ is a forest, seen as a posetal category, then the category $\widehat{X}$ of presheaves over~$X$ is arboreal with respect to the factorisation system (epimorphisms, monomorphisms). The paths are the representable functors along with the initial object, and the path-generation property amounts to the well-known fact that representables in a presheaf category are dense. See~\cite{reggio2023finitely} for more details.

This generalises item~\ref{i:forests-trees}: just observe that, if $X=\mathbb{N}$ is the poset of natural numbers with the usual total order, $\widehat{\mathbb{N}}$ is equivalent to the category of forests and their morphisms. Moreover, taking as $X$ a one-element poset, we see that the category of sets and functions is also a (trivial) example of arboreal category, in which the paths are the empty set and the singletons.
\end{enumerate}
\end{example}

\subsection{Games in arboreal categories}

\subsubsection{The functor \texorpdfstring{$\Path$}{P}}
The next result provides an analogue to the forgetful functor from the category $\EM(\G)$ of coalgebras for a game comonad $\G$ into the category of forests, and is a key tool in the theory of arboreal categories.

\begin{proposition}
Let $\A$ be an arboreal category. The assignment 
\[
(f\colon X\to Y) \ \longmapsto \ (\exists_{f}\colon \Path{X}\to \Path{Y})
\] 
yields a functor $\Path\colon \A\to\T$ into the category $\T$ of non-empty trees and forest morphisms.
\end{proposition}

\begin{example}
In the case of modal logic, $\Path\colon \EM(\Mk)\to \T$ is naturally isomorphic to the functor sending a synchronization tree to its underlying tree order. The same is true for $\EM(\Ek)$ and $\EM(\Pn)$, up to the equivalence between $\T$ and the category of forests (see Remark~\ref{rem:equiv-forest-tree}). The fact that each $\Path{X}$ is a \emph{non-empty} tree is due to the fact that arboreal categories have an initial object (obtained as an empty coproduct) which is a path; concretely, in the case of $\EM(\Ek)$ and $\EM(\Pn)$, this is the empty coalgebra. 
\end{example}

\subsubsection{The arboreal game \texorpdfstring{$\Ga$}{G}}
The tree order on $\Path X$ and the associated covering relation~$\cvr$ (cf.\ Footnote~\ref{footn:covering-rel}) allow us to define abstract games in any arboreal category~$\A$ by essentially playing on the corresponding trees.
Let $X,Y$ be any two objects of $\A$. We define a back-and-forth game 
\[
\Ga(X,Y)
\] 
played by Spoiler and Duplicator as follows:
\begin{itemize}\itemsep2pt
\item Positions in the game are pairs of path embeddings $(m,n)\in\Path{X}\times\Path{Y}$.
The winning relation $\W(X,Y)\subseteq \Path{X}\times\Path{Y}$ consists of the pairs $(m,n)$ such that $\dom(m)\cong\dom(n)$.
\item Let $\bot_X$ and $\bot_Y$ be the roots of $\Path{X}$ and $\Path{Y}$, respectively. If $(\bot_X, \bot_Y)\notin \W(X,Y)$, Duplicator loses the game. Otherwise, the initial position is $(\bot_X,\bot_Y)$.
\item At the start of each round, with current position $(m,n)\in\Path{X}\times\Path{Y}$, either Spoiler chooses some $m'\succ m$ and Duplicator must respond with some $n'\succ n$, or Spoiler chooses some $n''\succ n$ and Duplicator must respond with $m''\succ m$. 
\item Duplicator wins the round if they are able to respond and the new position is in $\W(X,Y)$. Duplicator wins the game if they have a strategy that is winning after~$t$ rounds, for all $t\geq 0$.
\end{itemize}

The game $\Ga(X,Y)$ is an abstraction of the back-and-forth games introduced in \S\ref{s:games}. For example, when $\A=\EM(\EkI)$, $X=\FI(A)$ and $Y=\FI(B)$ with $A,B$ $\sg$-structures, the game $\Ga(X,Y)$ is equivalent to the $k$-round \EF~game played between~$A$ and~$B$. Similarly for $\EM(\PnI)$ and $\EM(\Mk)$.

In the concrete setting, we have four variants of each game, depending on whether Spoiler can choose in both structures or only in one, and on whether the winning condition is given by the existence of a partial homomorphism or a partial isomorphism (or, in the case of modal logic, whether unary relations should only be preserved or also reflected). Accordingly, along with the basic back-and-forth games, which capture the full fragments $\LL$, we have their existential, positive, and existential positive variants, which capture $\exists\LL$, ${}^{+}\LL$, and $\exists^{+}\LL$, respectively.

Likewise, we can define the following variants of the game $\Ga(X,Y)$:
\begin{itemize}
\item In the \emph{existential game} \[\exists\Ga(X,Y)\] Spoiler plays only in $\Path{X}$ and Duplicator responds in $\Path{Y}$; the winning relation $\W(X,Y)$ is the same as in $\Ga(X,Y)$.
\item In the \emph{positive game} \[{}^{+}\Ga(X,Y)\] Spoiler can play in either $\Path{X}$ or $\Path{Y}$; the winning relation $\W^{+}(X,Y)\subseteq \Path{X}\times\Path{Y}$ consists of the pairs $(m,n)$ such that there exists a morphisms $\dom(m)\to\dom(n)$. 
\item In the \emph{existential positive game} \[\exists^{+}\Ga(X,Y)\] Spoiler plays only in $\Path{X}$ and Duplicator responds in $\Path{Y}$; the winning relation $\W^{+}(X,Y)$ is the same as in ${}^{+}\Ga(X,Y)$.
\end{itemize}

The following result relates the games $\Ga$, $\exists^{+}\Ga$ and $\exists\Ga$ to the corresponding relations induced by bisimulations, morphisms, and pathwise embeddings, respectively; the case of the game ${}^{+}\Ga$ is discussed below.

\begin{theorem}\label{t:arboreal-games-relations}
The following statements hold for all objects $X,Y$ of an arboreal category $\A$:
\begin{enumerate}
\item\label{i:prod-bf} Suppose $X$ and $Y$ admit a product in $\A$. Duplicator has a winning strategy in the game $\Ga(X,Y)$ if, and only if, $X$ and~$Y$ are bisimilar.
\item\label{i:P-faith-ep} Suppose the functor $\Path\colon\A\to\T$ is faithful. Duplicator has a winning strategy in the game $\exists^{+}\Ga(X,Y)$ if, and only if, there exists a morphism $X\to Y$.
\item\label{i:exist-pathwise-emb} Duplicator has a winning strategy in the game $\exists\Ga(X,Y)$ if, and only if, there exists a pathwise embedding $X\to Y$.
\end{enumerate}
\end{theorem}

\begin{remark}\label{rem:additional-assump}
We comment on the additional assumptions in items~\ref{i:prod-bf} and~\ref{i:P-faith-ep} of Theorem~\ref{t:arboreal-games-relations}. The existence of a product of $X$ and $Y$ in item~\ref{i:prod-bf} is needed to construct the vertex $W$ of a bisimulation between $X$ and $Y$ as an $\M$-subobject of $X\times Y$. 

Concerning item~\ref{i:P-faith-ep}, the following are equivalent for any arboreal category~$\A$:
\begin{enumerate}[label=(\alph*)]
\item The functor $\Path\colon\A\to\T$ is faithful.
\item\label{i:at-most-one-arrow} For any two paths $P,Q$ in $\A$, there exists at most one morphism $P\to Q$.
\end{enumerate}
A Duplicator winning strategy in $\exists^{+}\Ga(X,Y)$ induces a cocone with vertex~$Y$ over the diagram of path embeddings into $X$. The assumption that $\Path\colon\A\to\T$ be faithful ensures, by item~\ref{i:at-most-one-arrow}, that any diagram of paths commutes, and so this is a \emph{compatible} cocone which yields a mediating morphism $X\to Y$ because $X$ is path-generated. This additional assumption is not needed in item~\ref{i:exist-pathwise-emb}, because for any two paths $P,Q$ there is always at most one embedding $P\emb Q$ \cite[Lemma~3.15(c)]{AR2022}.
\end{remark}

The categories of coalgebras $\EM(\EkI)$, $\EM(\PnI)$ and $\EM(\Mk)$ satisfy all the assumptions in Theorem~\ref{t:arboreal-games-relations}, which can thus be regarded as an axiomatic generalisation of Theorems~\ref{th:ep-Kleisli-morphisms},~\ref{th:e-pathwise-emb} and~\ref{th:full-bisim}.

A similar result holds for the positive game ${}^{+}\Ga(X,Y)$, but requires extra care; see~\cite{ALR23}. First of all, since the concrete notion of positive bisimulation in Definition~\ref{d:positive-bisim} involves a set-theoretic bijection, we need to reformulate this notion slightly in the arboreal setting.
Given objects $X,Y$ of an arboreal category $\A$, we call \emph{positive bisimulation} from $X$ to~$Y$ a triple of arrows $(f,\iota,g)$
\[\begin{tikzcd}[]
W_{1} \arrow[r, "\iota"] \arrow[d,"f"', circlearrow] & W_{2} \arrow[d,"g", circlearrow] \\
X & Y
\end{tikzcd}\]
where $f$ and $g$ are open pathwise embeddings and $\Path{\iota}$ is a bijection. 

The existence of a Duplicator winning strategy in the game ${}^{+}\Ga(X,Y)$ is equivalent to the existence of a positive bisimulation from $X$ to $Y$, provided that $\A$ satisfies some further properties. In addition to requiring that $X$ and $Y$ admit a product, and that $\Path\colon\A\to\T$ be faithful, we require that each path be \emph{tree-connected}---a strong form of connectedness which is satisfied in the categories $\EM(\Ek)$, $\EM(\Pn)$ and $\EM(\Mk)$. The latter condition is related to the functor $\Path$ being a (Street) \emph{fibration} and implies that every arrow admits a decomposition of the form $g\circ \iota$ such that $\Path{\iota}$ is a bijection and~$g$ is a pathwise embedding, which is used to construct a positive bisimulation from~$X$ to~$Y$.\footnote{The notion of tree-connectedness is introduced in a forthcoming work by Tom\'a\v{s} Jakl and the second author, where it is shown that, under appropriate assumptions, the functor $\Path$ is a fibration and has a right adjoint.}

\section{Arboreal adjunctions}
\label{s:HPT}

In the examples, the bisimilarity relation available in the arboreal categories $\EM(\Ek)$, $\EM(\Pn)$ and $\EM(\Mk)$ is used to compare (pointed) relational structures $A,B$ by testing bisimilarity of the co-free coalgebras $F(A)$ and $F(B)$ or, in the presence of the equality symbol, of the coalgebras $\FI(A)$ and $\FI(B)$. In other words, we transfer the bisimilarity relation along a right adjoint functor. This leads to the following definition:

\begin{definition}
An \emph{arboreal adjunction} is an adjunction
\begin{equation}\label{eq:arboreal-adj}
\begin{tikzcd}[column sep=0.8em]
\E \arrow[hbend left]{rr}{F} & \text{\footnotesize{$\top$}} & \A \arrow[hbend left]{ll}{L}
\end{tikzcd}
\end{equation}
where $\A$ is an arboreal category and $\E$ is any category. We shall refer to $\E$ as the \emph{extensional category}; this is typically a category of (pointed) relational structures. 
\end{definition}

\begin{example}\label{ex:arboreal-adj}
The Eilenberg--Moore adjunction associated with any of the comonads $\Ek$, $\Pn$ and~$\Mk$ is arboreal.
The composite adjunctions as in eq.~\eqref{eq:I-adjunction} are also arboreal.
\end{example} 

\begin{remark}
\emph{Resource-indexed} arboreal adjunctions can also be defined, to take into account the resource parameters. 
\end{remark}

\subsection{\texorpdfstring{$F$-bisimilarity}{F-bisimilarity}}\label{s:F-bisimilarity}
Given an arboreal adjunction as in eq.~\eqref{eq:arboreal-adj}, the \mbox{\emph{$F$-bisimilarity}} relation~$\lr_{F}$ between objects $a,b$ of $\E$ is defined by
\[
a\lr_{F} b \ \text{ if, and only if, } \ Fa\lr Fb \ \text{ (i.e., $Fa$ and $Fb$ are bisimilar in $\A$).} 
\]
With this notation, item~\ref{i:prod-bf} of Theorem~\ref{t:arboreal-games-relations} entails that the equivalence relations $\equiv_{\FO_{k}}$, $\equiv_{L^{n}_{\infty,\omega}}$ and $\equiv_{\ML_{k}}$ between (pointed) structures coincide with $\lr_{F}$ for an appropriate choice of the right adjoint functor $F$. 
Similarly, we can transfer other relations from the arboreal category to the extensional one, yielding e.g.\ relations 
\[
\cong_{F} \ \text{ and } \ \to_{F}
\] 
between objects of $\E$.

One may ask if there is an \emph{upper bound} to the expressive power of the $F$-bisimilarity relation induced by an arboreal adjunction. Note that isomorphic objects are always $F$-bisimilar. Could we devise, for instance, an arboreal adjunction
\begin{equation}\label{eq:arb-adj-cs}
\begin{tikzcd}[column sep=0.8em]
\CS \arrow[hbend left]{rr}{F} & \text{\footnotesize{$\top$}} & \A \arrow[hbend left]{ll}{L}
\end{tikzcd}
\end{equation} 
such that $\lr_{F}$ coincides with isomorphism of structures? The next result provides an answer when the arboreal category $\A$ is \emph{locally finitely presentable} (in the sense of, e.g., \cite{AR1994,GU1971}) and the adjunction is \emph{finitely accessible}, i.e.\ $F$ preserves filtered colimits. Note that these assumptions are satisfied in the examples arising from game comonads: all arboreal categories in Example~\ref{ex:arboreal-cats} are locally finitely presentable and all adjunctions in Example~\ref{ex:arboreal-adj} are finitely accessible. 

\begin{theorem}[\cite{reggio2023finitely}]\label{th:expressive-power}
Given an arboreal adjunction as in eq.~\eqref{eq:arb-adj-cs}, suppose that $\A$ is locally finitely presentable and the adjunction is finitely accessible. Then ${\equiv_{L_{\infty,\omega}}}\subseteq {\lr_{F}}$. 
\end{theorem}

In other words, under the assumptions of the previous theorem, the $F$-bisimilarity relation cannot distinguish two structures that satisfy the same sentences in~$L_{\infty,\omega}$ or, equivalently, that are \emph{potentially isomorphic} \cite{Karp1965}. In fact, this bound is tight: there exists a finitely accessible arboreal adjunction such that ${\equiv_{L_{\infty,\omega}}} = {\lr_{F}}$, namely the one induced by the infinitary \EF~comonad $\Eo$ in Proposition~\ref{p:E-omega} (composed with the adjunction $H\dashv J$ between $\sg$-structures and $\sg^{I}$-structures).

\begin{remark}
Theorem~\ref{th:expressive-power} is a special case of a more general result which applies not only to extensional categories of the form $\CS$ but, more generally, to categories of models of \emph{cartesian theories} (see~\cite{coste76benabou} or \cite[\S D1.3]{Elephant2}). Furthermore, if we replace $F$-bisimilarity with \emph{$F$-back-and-forth equivalence}, which relates objects $a$ and $b$ exactly when Duplicator has a winning strategy in the game $\Ga(F(a),F(b))$, then the previous theorem can be extended to \emph{wooded categories}, a considerable weakening of the notion of arboreal category.\footnote{$F$-bisimilarity and $F$-back-and-forth equivalence coincide in any arboreal category with binary products by Theorem~\ref{t:arboreal-games-relations}, but need not coincide in wooded categories.}

For more details see~\cite{reggio2023finitely}, which is based on the use of Gabriel--Ulmer duality for locally finitely presentable categories~\cite{GU1971}, and on Hodges' \emph{word-constructions} \cite{hodges74,hodges75la}.
\end{remark}

\subsection{The bisimilar companion property and homomorphism preservation theorems}
\label{s:bcp-hpt}
The following property of arboreal adjunctions turns out to be a key \emph{dividing line}:
\begin{definition}
An arboreal adjunction $L\dashv F\colon \E\to \A$, with induced comonad $\G\coloneqq LF$, has the \emph{bisimilar companion property} if $a\lr_{F} \G(a)$ for all objects $a$ of $\E$.
\end{definition}

The arboreal adjunctions associated with the comonads $\Mk$ have the bisimilar companion property; in fact, the idempotency of the modal comonads implies the stronger property $a\cong_{F} \G(a)$. On the other hand, the arboreal adjunctions arising from the comonads $\Ek$ and $\Pn$ do not have the bisimilar companion property. In between, there are e.g.\ the \emph{guarded comonads}~\cite{Guarded2021}, corresponding to guarded fragments of first-order logic~\cite{HNvB1998}, which satisfy the bisimilar companion property but are not idempotent. 

The bisimilar companion property is correlated with the tractability of the corresponding logics: the modal and guarded fragments are decidable and have the tree-model property \cite{Vardi1997,Gradel2001}, while the quantifier rank and finite-variable fragments (which are ``expressive'', in the sense that they cover the whole of first-order logic) do not. These
observations provide a first step towards using structural properties of comonadic semantics to classify logic fragments and their expressive power.

From the comonadic perspective, the bisimilar companion property is central to the analysis of homomorphism preservation theorems in logic~\cite{AR2024}. Preservation theorems relate the syntactic shape of formulas to the semantic property of being preserved under a class of transformations. E.g.\ the classical \emph{homomorphism preservation theorem} (HPT), due to {\L}o\'s, Lyndon and Tarski~\cite{Los1955,Tarski1955,Lyndon1959}, states that a first-order sentence is preserved under homomorphisms between models (of a theory~$T$) if, and only if, it is equivalent (modulo~$T$) to an existential positive sentence. 
Rossman's \emph{Finite~HPT}~\cite{Rossman2008}, a seminal result in finite model theory, shows that the classical HPT holds relative to finite relational structures (with $T=\emptyset$ the empty theory). This is a striking result, as most preservation theorems fail on finite structures, see e.g.\ \cite{Tait1959,AG1987}. On his way to the celebrated Finite HPT, Rossman established the following \emph{Equirank HPT}, a remarkable resource-sensitive refinement of the classical result:

\begin{theorem}
Let $\phi$ be a first-order sentence of quantifier rank $\leq k$ in a relational signature $\sg$. Then $\phi$ is preserved under homomorphisms between $\sg$-structures if, and only if, it is equivalent to an existential positive sentence $\psi$ of quantifier rank $\leq k$.
\end{theorem}

To rephrase the previous result in a categorical language, we introduce the following terminology for an arbitrary full subcategory $\D$ of the extensional category $\E$:
\begin{itemize}
\item $\D$ is \emph{closed (in $\E$) under morphisms} if, whenever there is an arrow $a\to b$ in $\E$ with $a\in \D$, also $b\in \D$.
\item Given a relation $\nabla$ on the class of objects of $\E$, $\D$ is \emph{upwards closed (in $\E$) with respect to $\nabla$} if 
\[
\forall a,b \in \E, \text{ if } a\in\D \text{ and } a \mathbin{\nabla} b \text{ then } b\in\D.
\] 
\item If the previous condition is satisfied and $\nabla$ is an equivalence relation, we say that $\D$ is \emph{saturated under~$\nabla$}.
\end{itemize}  

For any arboreal adjunction $L\dashv F\colon \E\to \A$, we can then consider the following statement:
\begin{enumerate}[label=\textnormal{(HP)}]
\item\label{HP-abstract} For any full subcategory $\D$ of $\E$ saturated under $\lr_{F}$, $\D$ is closed under morphisms precisely when it is upwards closed with respect to $\to_{F}$.
\end{enumerate}
When considering the arboreal adjunction in eq.~\eqref{eq:I-adjunction} with $\G = \Ek$ the \EF~comonad, the full subcategories of $\CS$ saturated under $\lr_{F}$ coincide with the model classes of sentences in $\FO_{k}$, and those that are upwards closed with respect to $\to_{F}$ with the model classes of sentences in $\EPFO_{k}$. Therefore, in this case, property~\ref{HP-abstract} coincides with the Equirank HPT.

In general, property~\ref{HP-abstract} can be regarded as an abstract formulation of an \emph{equi-resource homomorphism preservation theorem}, stating that the complexity of a sentence~$\phi$ invariant under homomorphisms can be preserved when passing to an equivalent existential positive sentence~$\psi$. 

The right-to-left implication in~\ref{HP-abstract} is always satisfied and corresponds to the fact that existential positive sentences are preserved under homomorphisms. 
Consider now the reverse implication: let~$\D$ be a full subcategory of~$\E$ closed under morphisms and saturated under~$\lr_{F}$, and suppose that $a \to_{F} b$ for objects $a,b$ of~$\E$. By definition, this means that there is an arrow $F(a) \to F(b)$ and so, as $L$ is left adjoint to $F$, there is an arrow $\G(a) = LF(a) \to b$. If the arboreal adjunction satisfies the bisimilar companion property, we get
\begin{equation}\label{eq:HPT-bcp}
a \, \lr_{F}\, \G(a)\,  \to \, b.
\end{equation}
Therefore, $a\in \D$ implies $b\in \D$, and so $\D$ is upwards closed with respect to $\to_{F}$. That is,

\begin{proposition}\label{p:HPT-tame}
\ref{HP-abstract} holds for any arboreal adjunction with the bisimilar companion property.
\end{proposition}

The previous proposition yields, for example, equi-resource homomorphism preservation theorems for (graded) modal logic and guarded logic. Moreover, these results relativise to finite structures because the corresponding comonads restrict to finite structures; see \cite[\S\S5.1,~5.3]{AR2024}.

When an arboreal adjunction does \emph{not} have the bisimilar companion property, it can happen that $a \mathbin{\centernot\lr}_{F} \G(a)$ and so we cannot use eq.~\eqref{eq:HPT-bcp} to derive~\ref{HP-abstract}. We may try to ``force'' the bisimilar companion property by constructing companions $a^{*}\rl a$ and $\G(a)^{*}\rl \G(a)$ that are $F$-bisimilar: 
\[\begin{tikzcd}[column sep=0.1em]
a^* & {\lr_{F}} & \G(a)^* &&&&& &&&&& &&&&&  \\
a \arrow[xshift=-3pt]{u} \arrow[leftarrow,xshift=3pt]{u} & {\rl_{F}} & \G(a) \arrow[xshift=-3pt]{u} \arrow[leftarrow,xshift=3pt]{u} \arrow{rrrrrrrrrrrrrrr}{} &&&&& &&&&& &&&&&  b
\end{tikzcd}\] 
Since it is always the case that $a\rl_{F} \G(a)$ (i.e., $a\to_{F} \G(a)$ and $\G(a)\to_{F} a$), we are essentially upgrading the relation $\rl_{F}$ to the finer relation $\lr_{F}$. This is the strategy adopted by Rossman to prove his homomorphism preservation theorems; for example the Equirank HPT is based on an upgrading of the relation $\equiv_{\EPFO_{k}}$ to the relation $\equiv_{\FO_{k}}$. This kind of \emph{upgrading~argument} is a powerful tool in finite model theory for obtaining preservation theorems and, more generally, expressive completeness results; see~\cite{Otto2011} for an overview of this method.

A fully axiomatic proof of the Equirank HPT, in the language of arboreal adjunctions, was given in~\cite{AR2024} using the argument outlined above. The main hurdle is to construct the companions~$a^{*}$ and~$\G(a)^{*}$, which are called \emph{extendable covers} of~$a$ and~$\G(a)$ in \emph{op.\ cit.} These are defined as colimits of certain countable chains of sections in the extensional category, similar to a small object argument in abstract homotopy theory; we come back to this topic in the next section. 

Finally, let us mention that it is an open problem whether a Finite Equirank HPT holds. On the other hand, by combining the classical (respectively, Finite) HPT with the main combinatorial construction in~\cite{BC2019}, one obtains a \emph{(Finite) Equivariable HPT}. An axiomatic understanding of the latter, as well as of the Finite HPT, has so far been elusive; partial results were obtained in~\cite{Paine2020}.

\section{Towards a homotopical view on logical resources}
\label{s:homotopical}

Many questions in (finite) model theory, and more generally in logic, can be phrased as \emph{model existence} problems: is there a (finite) model satisfying a collection of logically specified properties? In classical model theory, the Compactness Theorem is a basic tool to tackle these problems. On the other hand, in finite and resource-sensitive model theory one typically addresses a model existence problem via the \emph{model construction} method, which consists in explicitly constructing a model satisfying the desired properties. We saw an example of this in \S\ref{s:bcp-hpt}, where an upgrading argument is combined with the construction of extendable covers of certain relational structures. 

In general, an object obtained by means of a model construction argument is overdetermined, in the sense that it is specified up to isomorphism and not up to equivalence with respect to the prescribed logical properties.
This idea is pervasive in mathematics and computer science: we may not want to distinguish between two topological spaces that are homotopy equivalent, or two programs that on the same input give the same output.

Our aim in this section is to make the analogy between resource-sensitive model theory and homotopy more precise using the language of abstract homotopy, also called \emph{categorical homotopy}. We posit that categorical homotopy provides a flexible setting in which to manipulate objects up to logical equivalence and perform constructions that are invariant under such equivalences; this idea is illustrated in the case of modal logic in \S\ref{s:model-modal}.

\subsection{Quillen model categories}

The extremely influential concept of \emph{model category} was introduced by Daniel Quillen to provide an abstract setting for homotopy theory~\cite{Quillen1967}. Broadly speaking, a model category is a category equipped with a notion of weak equivalences $\We$ between objects, along with classes $\Cof$ and $\Fib$ of ``good'' injections and surjections, respectively.

\begin{definition}
A \emph{(Quillen) model structure} on a (locally small, finitely bicomplete) category $\D$ is a triple $(\We,\Cof,\Fib)$ of classes of morphisms in $\D$ satisfying the following two properties:
\begin{enumerate}[label=(\roman*)]
\item The class $\We$ has the 2-out-of-3 property: given composable morphisms $u\colon X\to Y$ and $v\colon Y\to Z$, if any two of $u, v$, and $v\circ u$ are in $\We$, so is the third.
\item $(\Cof\cap \We,\Fib)$ and $(\Cof,\Fib\cap\We)$ are weak factorisation systems (cf.\ Definition~\ref{def:f-s}).
\end{enumerate} 
A \emph{(Quillen) model category} is a category $\D$ equipped with a model structure $(\We,\Cof,\Fib)$.
\end{definition}
We refer to the arrows in $\We$, $\Cof$ and $\Fib$ as \emph{weak equivalences}, \emph{cofibrations} and \emph{fibrations}, respectively. The arrows in $\Cof\cap \We$ are called \emph{trivial cofibrations}, and those in $\Fib\cap\We$ \emph{trivial fibrations}. 
Cofibrations and fibrations are denoted, respectively, by $\cof$ and $\fib$, and weak equivalences by $\we$. E.g., $\tfib$ denotes a trivial fibration.

\begin{remark}\label{rem:overdetermined}
The data defining a model structure is overdetermined: any two of the classes $\We$, $\Cof$ and $\Fib$ determine the third. For example, an arrow is a weak equivalence exactly when it can be factored as a trivial cofibration followed by a trivial fibration.
In fact, by a result of Joyal (see \cite[Proposition~E.1.10]{Joyal2008} or \cite[Theorem 15.3.1]{Riehl2014}), a model structure on a category is completely determined by its cofibrations together with its class of \emph{fibrant} objects:
\end{remark}

\begin{definition}
An object $X$ of a model category is \emph{fibrant} if the unique morphism $X\to\one$ to the terminal object is a fibration.
\end{definition}

A convenient notion of morphism between model categories is that of Quillen adjunction:
\begin{definition}
A \emph{Quillen adjunction} between model categories $\D$ and $\D'$ is an adjoint pair of functors
\[\begin{tikzcd}[column sep=1.0em]
\D \arrow[bend left=35]{rr}[description]{G} & \text{\tiny{$\top$}} & \D' \arrow[bend left=35]{ll}[description]{F}
\end{tikzcd}\]
such that $F$ preserves cofibrations and $G$ preserves fibrations. 
\end{definition}

A special class of model categories that we will encounter in the context of modal logic is that of Cisinski model categories~\cite{Cisinski2006}. 
Recall that a model structure $(\We,\Cof,\Fib)$ on a Grothendieck topos is a \emph{Cisinski model structure} if it satisfies the following properties:
\begin{enumerate}[label=(\roman*)]
\item it is \emph{cofibrantly generated}, i.e.\ there exist (small) sets $I,J$ permitting the small object argument\footnote{See e.g.\ \cite[\S 12.2]{Riehl2014} for a discussion of what it means to ``permit the small object argument''.} such that $\Cof = \llp{(\rlp{I})}$ and $\Cof \cap \We = \llp{(\rlp{J})}$.
\item $\Cof$ coincides with the class of monomorphisms.
\end{enumerate}
By Remark~\ref{rem:overdetermined}, a Cisinski model structure is completely determined by its class of fibrant objects.

\subsection{A model category for modal logic}\label{s:model-modal}
Fix an arbitrary positive integer $k$ and recall from \S\ref{s:coalgebras-modal} the category $\EM(\Mk)$ of Eilenberg--Moore coalgebras for the modal comonad $\Mk$. We denote by 
\[
\St_k\] 
its (wide) subcategory having the same objects but morphisms the pathwise embeddings. Concretely,~$\St_k$ can be thought of as the category of synchronization trees of height $\leq k$ and homomorphisms of Kripke models that reflect the unary relations (cf.\ Theorem~\ref{t:coalgebras-Mk}).

The category $\St_k$ is typically neither complete nor cocomplete. In fact, if the modal vocabulary contains at least one unary relation symbol, then $\St_k$ admits neither an initial object nor a terminal object. We can (co)complete it by embedding it into a presheaf category: if $\P_{k}$ denotes the full subcategory of $\St_{k}$ defined by the paths (i.e., the finite branches), the restricted Yoneda embedding
\begin{equation}\label{eq:Yoneda-S-k}
\St_k \to \w{\P_k}, \ \ \ A \mapsto \left(\St_k( - , A)\colon \P_k^\op \to \Set \right)
\end{equation}
is full and faithful because $\P_{k}$ is dense in $\St_{k}$.
We can thus identify $\St_k$ (as well as $\P_k$) with a full subcategory of $\w{\P_k}$. 
Note that the category $\P_{k}$ is a preorder; for simplicity, we shall work with a small skeleton of the class of paths, so that $\P_{k}$ can be identified with a (small) forest.

We argue that the presheaf categories $\w{\P_k}$ offer a setting in which some key notions and constructions in modal logic admit a natural homotopical interpretation.

\begin{proposition}\label{p:model-structure}
The category $\w{\P_k}$ admits a Cisinski model structure in which an object $X$ is fibrant if, and only if, it satisfies the following property: For all arrows $i\colon P\to Q$ in~$\P_{k}$, if there is an arrow $Q\to X$ then every arrow $P\to X$ admits a lifting along $i$:
\[\begin{tikzcd}
P \arrow{d}[swap]{i} \arrow{r} & X \\
Q \arrow[dashed]{ur} & {}
\end{tikzcd}\]
\end{proposition}

What is more, the stratification by depth of modal formulas induces a tower of Quillen adjunctions
\[  \begin{tikzcd}[column sep = 3em]
\widehat{\P_1} \arrow[bend left=50,leftarrow,yshift=-3pt]{r} \arrow[r, hookrightarrow, "\top" {yshift = 3pt}] & \widehat{\P_2}  \arrow[bend left=50,leftarrow,yshift=-3pt]{r} \arrow[r, hookrightarrow, "\top" {yshift = 3pt}] & \widehat{\P_3} \arrow[bend left=50,leftarrow,yshift=-3pt]{r} \arrow[r, hookrightarrow, "\top" {yshift = 3pt}] & \widehat{\P_4} \arrow[r, draw=none, "\hspace{-1em} \ \ \ \cdots" description] & {}
\end{tikzcd} \]

Restricting this model structure to the subcategory $\St_{k}$, we recover familiar classes of Kripke homomorphisms. In fact, an arrow $f\colon A\to B$ in $\St_k$ is a
\begin{itemize}
\item cofibration if, and only if, it is an embedding of Kripke models, and a
\item trivial fibration if, and only if, it is a (surjective) \emph{p-morphism} \cite[Definition~2.10]{blackburn2002modal}.
\end{itemize}

We show how to use this homotopical structure to give a simple interpretation of the following preservation result:
\begin{theorem}\label{th:existential-hpt}
A modal formula of depth $\leq k$ is preserved under extensions of Kripke models if, and only if, it is equivalent to an existential modal formula of depth $\leq k$. Further, the result relativises to finite Kripke models.
\end{theorem}
The first part of the previous result, without reference to the modal depth of formulas, was first proved by van Benthem (cf.~\cite{HNvB1998}) and provides a semantic characterisation of the existential fragment $\EML$ of modal logic. As such, it is akin to the classical {\L}o\'s--Tarski preservation theorem for first-order logic~\cite{Los1955,Tarski1955}, which states that a first-order sentence is preserved under extensions of models if, and only if, it is equivalent to an existential sentence.
The resource-sensitive refinement, which takes into account the modal depth of formulas, and the relativisation to finite Kripke models are due to Rosen~\cite{Rosen1997}. 

In the same spirit of the categorical account of homomorphism preservation theorems in \S\ref{s:bcp-hpt}, the first part of Theorem~\ref{th:existential-hpt} can be reformulated as follows:
\begin{enumerate}[label=\textnormal{(EP)}]
\item\label{EP-modal} 
For any full subcategory $\D$ of $\CSstar$ saturated under $\equiv_{\ML_{k}}$, $\D$ is closed under embeddings precisely when it is upwards closed with respect to $\IMP_{\EML_{k}}$.
\end{enumerate}
Property~\ref{EP-modal} can be proved by an upgrading argument, which upgrades the relation $\IMP_{\EML_{k}}$ to the relation $\equiv_{\ML_{k}}$. For the non-trivial direction in~\ref{EP-modal}, suppose that $\D$ is a full subcategory of $\CSstar$ saturated under $\equiv_{\ML_{k}}$ and closed under embeddings. If $(A,a),(B,b)\in \D$ satisfy 
\[
(A,a)\IMP_{\EML_{k}} (B,b)
\] 
then, by item~\ref{i:exist-pe-modal} in Theorem~\ref{th:e-pathwise-emb}, there is an arrow $f\colon F(A,a)\to F(B,b)$ in $\St_{k}$ from the \mbox{$k$-unravelling} of $(A,a)$ to that of $(B,b)$. To conclude that $\D$ is upwards closed with respect to $\IMP_{\EML_{k}}$, it suffices to find a Kripke model $X$ as displayed below (note that we use the bisimilar companion property).
\[\begin{tikzcd}[column sep=-0.5em]
&& X & {\equiv_{\ML_{k}}} & F(B,b) & & \\
(A,a)&{\equiv_{\ML_{k}}} & F(A,a) \arrow[rightarrowtail]{u} \arrow{rr}{f} &  & F(B,b) \arrow[equal]{u} & {\equiv_{\ML_{k}}} & (B,b)
\end{tikzcd}\] 

In turn, such an $X$ can be obtained by taking an appropriate decomposition of $f$ as a cofibration followed by a trivial fibration (any arrow in a model category admits such a decomposition, but we need to ensure that $X$ is a Kripke model and not an arbitrary presheaf): 
\begin{lemma}\label{l:S-k-closed-under-fact}
Any morphism $(A_{1},a_{1})\to (A_{2},a_{2})$ in $\St_k$ can be factored as a cofibration followed by a trivial fibration
\[
(A_{1},a_{1}) \cof X \tfib (A_{2},a_{2})
\] 
with $X\in \St_k$. Moreover, if $A_{1}$ and $A_{2}$ are finite Kripke models, we can assume that $X$ is finite.
\end{lemma}

We have seen above that cofibrations and trivial fibrations between objects of $\St_{k}$ coincide with embeddings and p-morphisms, respectively. Since the latter preserve and reflect \emph{all} modal formulas, the Kripke model $X$ provided by Lemma~\ref{l:S-k-closed-under-fact} satisfies the upgrading argument.
This proves the first part of Theorem~\ref{th:existential-hpt}; the relativisation to finite models follows from the fact that we can take $X$ to be finite if the $k$-unravellings $F(A,a)$ and $F(B,b)$ are finite, which is the case if $A$ and $B$ are finite.

The characterisation of p-morphisms between objects of $\St_{k}$ as the trivial fibrations in the model structure on $\w{\P_{k}}$ suggests the following definition.

\begin{definition}
A \emph{Morita equivalence} between objects $X$ and $Y$ of a model category is a span of trivial fibrations connecting the two objects:
\[ 
X \tfibleft Z \tfib Y
\]
We say that $X$ and $Y$ are \emph{Morita equivalent} if there exists a Morita equivalence between them.\footnote{The terminology ``Morita equivalence'' arises from the theory of Lie groupoids (see e.g.~\cite{Zhu2009}), and is used with a different meaning in algebra (for rings inducing equivalent categories of modules) and categorical logic (to refer to theories with equivalent categories of models in any topos).}
\end{definition}

\begin{remark}
While Morita equivalent objects are clearly weakly equivalent, the converse need not hold.
On the other hand, in view of Kenneth Brown’s Factorization lemma \cite[p.~421]{Brown1973}, combined with Whitehead's theorem (see e.g.\ \cite[Theorem~7.5.10]{Hirschhorn2003}), Morita equivalence between fibrant objects coincides with homotopy equivalence. 
\end{remark}

Given $k$-unravellings $F(A,a)$ and $F(B,b)$ of pointed Kripke models $(A,a)$ and $(B,b)$, respectively, any bisimulation between $F(A,a)$ and $F(B,b)$ in the arboreal category $\EM(\Mk)$ induces a Morita equivalence between $F(A,a)$ and $F(B,b)$ in the model category $\w{\P_{k}}$. In general, there are more Morita equivalences than bisimulations, since \emph{every} presheaf in $\w{\P_{k}}$ can in principle be the vertex of a Morita equivalence. However, it turns out that $F(A,a)$ and $F(B,b)$ are bisimilar in $\EM(\Mk)$ if, and only if, they are Morita equivalent in $\w{\P_{k}}$. Thus, by item~\ref{i:modal-b-and-f} in Theorem~\ref{th:full-bisim},
\[
(A,a)\equiv_{\ML_{k}} (B,b) \ \Longleftrightarrow \ \text{ $F(A,a)$ and $F(B,b)$ are Morita equivalent in $\w{\P_{k}}$}.
\]
Moreover, Morita equivalence in $\w{\P_{\omega}}$, where $\P_{\omega}= \bigcup_{k<\omega}{\P_{k}}$, captures precisely bisimilarity of Kripke models, in the usual sense of modal logic.

This observation allows, for example, to study Hennessy--Milner classes of Kripke models via \emph{presheaves of Morita equivalences}; for more details, we refer the interested reader to~\cite{Reggio2023modal}.

\subsection{Through the lens of modal logic: presheaves over forests}
It is well known that presheaves over posets are an appropriate setting for modal logic, so the occurrence of the presheaf categories $\w{\P_{k}}$ in the homotopy interpretation of (bounded depth) modal logic outlined in \S\ref{s:model-modal} is perhaps not surprising.

However, it is worth observing that the perspective of game comonads and arboreal categories points to a modal-like view on many fragments of (infinitary) first-order logic and corresponding decompositions of structures. A similar idea was put forward by Otto in~\cite{Otto2011} through the study of modal translations induced by \emph{observable configurations} of relational structures. We outline two connections between arboreal categories and modal logic which may be important for extending the homotopical approach to other fragments of first-order logic.

The first connection stems from a simple rephrasing of item~\ref{i:prod-bf} in Theorem~\ref{t:arboreal-games-relations}. Fix an arboreal category $\A$ with binary products; we shall assume that $\A$ contains, up to isomorphism, only a set~$\Gamma$ of paths. Consider the modal vocabulary whose set of unary relation symbols is~$\Gamma$ and whose only binary relation symbol is denoted by~$\acc$. For each object $X$ of $\A$, the tree $\Path{X}$ can be regarded as a Kripke model for this vocabulary, where~$\acc$ is interpreted as the covering relation~$\prec$ associated with the tree order, and the interpretation of $P\in \Gamma$ consists of the (equivalence classes of) path embeddings $m\in \Path{X}$ whose domain is isomorphic to $P$.
We write $\Pathast{X}$ for the ensuing Kripke model, whose distinguished element is the root.
With this notation, item~\ref{i:prod-bf} in Theorem~\ref{t:arboreal-games-relations} states that:
\begin{theorem}
$X,Y\in \A$ are bisimilar if, and only if, $\Pathast{X}$ and $\Pathast{Y}$ are bisimilar as Kripke models.
\end{theorem}
When combined with the $F$-bisimilarity relation induced by an arboreal adjunction (see \S\ref{s:F-bisimilarity}), the previous result can be regarded as a categorification of Otto's slogan ``Bisimulation as the master game''~\cite[\S6.2.4]{Otto2011}. For example, in the case of the pebbling comonad, it implies that structures $A,B$ satisfy the same sentences in $n$-variable infinitary logic $L^{n}_{\infty,\omega}$ if, and only if, the corresponding tree-ordered Kripke models are bisimilar. If $A$ and $B$ are finite structures, a similar statement holds with $\FO^{n}$ instead of $L^{n}_{\infty,\omega}$, cf.\ Corollary~\ref{cor:pebble-fin-struct}, but the trees are always infinite.

The second connection arises from considering presheaves over categories of paths. The arboreal category $\A$ embeds into the presheaf category over its full subcategory of paths (cf.\ Footnote~\ref{footn:dense-paths}). The full subcategory of paths is a preorder if, and only if, the functor $\Path\colon \A\to\T$ is faithful.
On the other hand, the subcategory $\A_{p}$ of $\A$ defined by the paths and the \emph{embeddings} between them is always a preorder and in fact, upon fixing a small skeleton, it is a forest order; cf.\ Remark~\ref{rem:additional-assump}. 
The restricted Yoneda embedding gives a functor
\[
\mathcal{I}\colon \A \to \w{\A_p}, \ \ \ X \mapsto \left(\A( - , X)\colon \A_p^\op \to \Set \right)
\] 
which is not faithful in general, unless we restrict it to the wide subcategory of $\A$ defined by the pathwise embeddings, cf.\ eq.~\eqref{eq:Yoneda-S-k}. Yet, the category $\w{\A_p}$ is arboreal by item~\ref{i:presh-forest} in Example~\ref{ex:arboreal-cats}, and the functor $\mathcal{I}$ preserves and reflects the bisimilarity relation.\footnote{This (as yet unpublished) result was obtained by the second-named author in collaboration with Colin Riba.}
This observation is central to the study of Morita equivalence in the presheaf categories $\w{\P_{k}}$ in the modal case, and suggests the possibility of extending the homotopical analysis beyond modal logic. 

Let us conclude by pointing out that objects of $\w{\P_{k}}$ can be regarded as generalised Kripke models (for simplicity, we assume that the modal language contains a single binary relation symbol~$\acc$):
\begin{definition}\label{def:generalised-Kripke-model}
A \emph{generalised Kripke model structure} on a presheaf $X\in\w{\P_k}$ is given by:
\begin{itemize}
\item an object $\ud{P}\in \P_k$ and a distinguished element $\ud{x}\in X(\ud{P})$, and
\item for each unary relation symbol $S$, a set $\{S_Q^{X} \subseteq X(Q) \mid Q\in\P_k\}$.
\end{itemize}
For $x_1\in X(Q_1)$ and $x_2\in X(Q_2)$, $x_1 \acc x_2$ if and only if
\begin{enumerate}[label=(\roman*)]
\item $Q_1 \prec Q_2$ in the forest order of $\P_k$, and
\item the unique embedding $i\colon Q_1\emb Q_2$ satisfies $X(i)\colon x_2\mapsto x_1$.
\end{enumerate}

We denote this by $(X,\ud{P},\ud{x})$, or even just $(X,\ud{x})$, and refer to it as a \emph{generalised Kripke model}. 
\end{definition}

Each $A\in \St_k$ (identified with the corresponding representable presheaf) can be equipped with a canonical structure of generalised Kripke model. Recall that, for all $Q\in \P_k$, the elements of $A(Q)$ are the embeddings $Q\emb A$. In turn, each such embedding determines a unique element of $A$, namely the image of the top element of $Q$. 
We let $\ud{P}$ be the unique root of $\P_k$ such that there is a (unique) embedding $m\colon \ud{P}\emb A$, and let $\ud{x}=m\in A(\ud{P})$. For each unary relation symbol $S$ and object $Q\in \P_k$, we take as $S_Q^{A}$ the set of embeddings $Q\emb A$ whose corresponding element of $A$ satisfies $S$. 
Finally, under the correspondence between elements of $A$ and embeddings $Q\emb A$ (with $Q\in \P_k$), the relation $\acc$ in Definition~\ref{def:generalised-Kripke-model} coincides with the accessibility relation of $A$.

We can then extend in a straightforward manner the satisfaction relation for modal formulas from Kripke models to generalised Kripke models. Alternatively, generalised Kripke models can be studied using the usual Kripke semantics by considering their \emph{externalisation}: this essentially amounts to constructing a transition system from a presheaf using the Grothendieck construction.
\begin{definition}
The \emph{externalisation} of a generalised Kripke model $(X,\ud{x})$ is the Kripke model $\Ext(X,\ud{x})$ defined as follows: the elements of $\Ext(X,\ud{x})$ are the tuples $(Q,y)$ such that $y\in X(Q)$, and the distinguished element is $(\ud{P},\ud{x})$. The accessibility relation $\acc$ is defined as in Definition~\ref{def:generalised-Kripke-model}, and similarly for the unary relation symbols (i.e., $(Q,y)$ satisfies $S$ if $y\in S_Q^{X}$).
\end{definition} 

For all generalised Kripke models $(X,\ud{x})$ and all modal formulas $\phi$, we have
\[
X,\ud{x}\models \phi \ \Longleftrightarrow \ \Ext(X,\ud{x}) \models \phi.
\]
Moreover, the assignment $(X,\ud{x})\mapsto \Ext(X,\ud{x})$ is functorial with respect to the obvious notion of morphism of generalised Kripke models (which are, in particular, natural transformations). The externalisation functor allows us to give a logic (or model-theoretic) meaning to the homotopical structure on $\w{\P_k}$. In fact, if $f$ is a morphism of generalised Kripke models that reflects the unary relations, then (the underlying natural transformation of) $f$ is a
\begin{itemize}
\item cofibration if, and only if, $\Ext(f)$ is an embedding of Kripke models, and a
\item trivial fibration if, and only if, $\Ext(f)$ is a p-morphism.
\end{itemize}

If we take the presheaf category $\w{\A_{p}}$, where $\A$ is the arboreal category corresponding to, say, $n$-variable logic or first-order logic with bounded quantifier rank, the approach outlined above yields a \emph{translation} $(\ \ )^{*}$ of these logic fragments into modal logic (for the category $\w{\P_k}$, this translation is the identity). The properties of this translation remain to be fully understood, and may allow one to extend the techniques described in \S\ref{s:model-modal} to first-order logic. Consider for example the following 

\vspace{0.5em}
\noindent\emph{Open problem.} Does an Equirank {\L}o\'s--Tarski theorem hold? 

\vspace{0.5em}
\noindent Working with the arboreal category corresponding to the \EF~comonad, a positive answer to this question would follow if we could prove that for every first-order sentence~$\phi$ that is preserved under embeddings, its modal translation $\phi^{*}$ is also preserved under embeddings.

\bibliographystyle{ACM-Reference-Format-Journals}


\end{document}